\documentclass[12pt]{article}
\usepackage{amssymb}
\usepackage{amsmath}
\usepackage{euscript}
\usepackage{bbold}
\usepackage{bm}

\numberwithin{equation}{section}       

\newcommand{\be}{\begin{equation}}
\newcommand{\ee}{\end{equation}}
\newcommand{\bea}{\begin{eqnarray}}
\newcommand{\eea}{\end{eqnarray}}
\newcommand{\nn}{\nonumber \\}
\newcommand{\reef}[1]{(\ref{#1})}
\newcommand{\wed}{\wedge }

\newcommand{\e}{\epsilon}  
\newcommand{\trsp}{{\mathrm T}}

\newcommand{\ex}{{\mathrm e}}
\newcommand{\diff}{{\mathrm d}}
\newcommand{\de}{\partial}

\newcommand{\bbR}{\mathbb{R}}
\newcommand{\bbH}{\mathbb{H}}

\newcommand{\bbT}{\mathbb{T}}
\newcommand{\bbS}{\mathbb{S}}
\newcommand{\id}{\mathbb{1}}

\newcommand{\rep}[1]{$\mathbf{#1}$}
\newcommand{\vol}{\mathrm{vol}}
\newcommand{\con}{\lrcorner\,}

\begin{document}


\begin{titlepage}

\vfill

\begin{center}

\baselineskip=16pt

\begin{flushright}
QMUL-PH-03-07\\
hep-th/0306225\\
\end{flushright}

\vfill

{\Large\bf $G$-Structures, Fluxes and Calibrations\\
\vskip 4mm
in M-Theory}

\vskip 1cm 

Dario Martelli and James Sparks

\vskip 1cm

\textsl{Department of Physics\\ Queen Mary, University of London\\
  Mile End Rd, London E1 4NS, U.K.}  

\vskip 0.5cm

{\tt d.martelli@qmul.ac.uk}~, $\quad$  {\tt j.sparks@qmul.ac.uk}

\end{center}

\vfill

\begin{center}
\textbf{Abstract}
\end{center}

\begin{quote}

We study the most general supersymmetric warped M-theory backgrounds with 
non-trivial $G$-flux of the type $\bbR^{1,2}\times M_8$ 
and AdS$_3\times M_8$. We give a set of 
necessary and sufficient conditions for preservation of supersymmetry 
which are phrased in terms of $G$-structures and their intrinsic torsion.
These equations may be interpreted as calibration conditions for a 
static ``dyonic'' M-brane, that is, an M5-brane with self-dual three-form 
turned on. When the electric flux is turned 
off we obtain the supersymmetry conditions and non-linear PDEs describing
 M5-branes wrapped on associative and special Lagrangian three-cycles in
manifolds with $G_2$ and $SU(3)$ structures, respectively. 
As an illustration of our formalism, we recover the 1/2-BPS dyonic M-brane, 
and also construct some new examples.

\end{quote}

\vfill

\end{titlepage}


\section{Introduction}

Recently there has been considerable interest in trying to understand
the types of geometries that arise in supersymmetric solutions of 
supergravity theories. When all fields are turned off, apart from the 
metric, it has long been known that supersymmetric solutions are described
by special holonomy manifolds -- for example, Calabi-Yau manifolds or 
manifolds of $G_2$ holonomy. 
However, for many applications one is interested in solutions where the fluxes 
are turned on. These include important areas of research, such as the AdS/CFT
correspondence, or phenomenological models based on string/M-theory 
compactifications.

Until recently, the study of supersymmetric solutions with non-vanishing fluxes
has been based mostly on physically motivated ansatze for the supergravity 
Killing spinor equations. While this method has led to many interesting
results, a more systematic approach is clearly desirable. In \cite{gmpw}
it was advocated that the $G$-structures defined by the Killing spinors 
provide such a formalism. 
Subsequent works have used this approach to analyse and classify 
supersymmetric backgrounds in various
supergravity theories \cite{d5,dan,dallagata,GP,KMPT,intrinsic,ALE,GG,d6}.
Using the language 
of $G$-structures and their ``intrinsic torsion''
one can rewrite the supersymmetry equations of interest in terms of an
equivalent set of first-order equations for a particular set of forms.

Another point, emphasized in \cite{gmpw} (and based on \cite{gkmw}), is the 
fact 
that some of the resulting conditions have an interpretation in terms of 
``generalised calibrations'' \cite{jan1,jan2}. 
This was further elaborated on in \cite{GP} and 
 \cite{intrinsic}. Generalised calibrations extend to backgrounds with 
fluxes the original notion of calibrations in special holonomy manifolds
\cite{HL}, and
their physical significance is then that supersymmetric probe 
branes have minimal {\em energy}.  On a more practical level, the formalism 
based on $G$-structures 
can often be very useful for actually finding new solutions
in a given supergravity theory. For instance, in  \cite{gmpw,GP,intrinsic}
new examples were found this way, while in lower dimensions 
\cite{d5,GG,d6} the general form for {\em all} supersymmetric
solutions was given.

In this paper we study M-theory on eight-manifolds -- that is, supersymmetric
warped M-theory backgrounds of the type $M_3\times M_8$, with $M_3$ either 
Minkowski$_3$ or AdS$_3$ space. 
Supersymmetric compactifications of M-theory to three dimensions have been 
considered
before in \cite{Becker1,sethi,Gukov:1999ya,Becker2,GS,AOG}. 
The types of geometries 
described in these papers may be thought of as M2-brane solutions where 
the transverse space is a manifold of special holonomy.
Alternatively, one may think of them as compactifications on a special 
holonomy manifold where 
one includes some number of space-filling M2-branes in the vacuum.

One of our motivations was to investigate 
more general types of supersymmetric solutions to M-theory on eight-manifolds.
In particular, there should clearly be another way to obtain an 
$\mathcal{N}=1$ Minkowski vacuum from M-theory -- namely, 
one may wrap M5-branes over a supersymmetric three-cycle in a $G_2$-holonomy
manifold (times an $S^1$). After including the  backreaction of the
M5-brane on the geometry, one no longer expects the eight-manifold
to have special holonomy, but rather a more general $G_2$-structure 
with intrinsic torsion related to the $G$-flux. Similarly, M5-branes
wrapped on special Lagrangian three-cycles in a Calabi-Yau three-fold 
yield ${\cal N}=2$ in three dimensions. 
We will show how these various geometries may be obtained by relaxing the 
assumptions
of \cite{Becker1,Becker2}; in particular we relax the assumption that the 
internal spinor is chiral. Furthermore, this  generalisation yields
supersymmetric AdS$_3$ compactifications, which were excluded before.
The method we use relies on local equations, and thus also covers non-compact 
geometries; examples of typical interest are solutions
describing wrapped branes or brane intersections.

The M-theory five-brane has a self-dual three-form gauge field that 
propagates on its world-volume. Turning on this field induces an 
electric coupling to the $C$-field, and therefore also an M2-brane charge.
Thus the backreaction of such a ``dyonic'' M5-brane should correspond to  
some more general supersymmetric solution with electric and magnetic $G$-flux.
In fact, we will see how such solutions arise in our formalism.
One can argue that the most general supersymmetric solution of the form 
$M_3\times M_8$ is of 
this type, with the M2-brane solutions being a limit in which the M5-brane 
disappears completely.

The plan of the paper is as follows. In section \ref{oldstuff} we give a 
brief summary of what is known about M-theory on eight-manifolds. This
will also allow us to introduce our notations and conventions. We then 
describe how one extends
the analysis to allow for more general supersymmetric solutions with
fluxes. The key point is to allow for a generic spinor on the internal space 
-- in particular, we do not impose that it be chiral. Thus, in addition 
to the M2-brane-type of solutions, one also expects
M5-brane-type solutions, including ``dyonic'' or ``interpolating'' solutions 
which have both charges present, and also AdS$_3$ solutions.

In section \ref{bilinears} we show how the conditions for supersymmetry may
be recast into the language of $G$-structures and intrinsic torsion. 
In particular, we argue that there is a $G_2 \subset SO(8)$ structure and 
obtain a simple set of differential conditions on the forms that comprise it.
By examining the intrinsic torsion one can show that these conditions 
are necessary and sufficient for supersymmetry. 
We also give the Bianchi identity and equations of motion
in this formalism and briefly discuss the issue of compact eight-manifolds. 
When the external manifold is $\bbR^{1,2}$, 
a simple inspection of the Einstein equations shows that 
one cannot have compact manifolds with flux, unless higher order corrections
are included.

In section \ref{cal} we turn our attention to the physical interpretation of
the differential condtions on the $G_2$-structure.
We show how these may be interpreted as generalised 
calibration conditions for the M5-brane. We argue that the geometries
that these equations describe correspond to ``dyonic'' M5-branes wrapped
over associative three-cycles in a $G_2$-holonomy manifold. Moreover, we
show that supersymmetric probe M5-branes saturate a calibration bound on their 
energy. We find that the M5-brane world-volume theory gives rise
not only to an M5-brane type of calibration, but also one gets
the M2-brane calibration ``for free''.

In section \ref{puremagnetic} we specialise our discussion to the case of 
``pure'' M5-branes (that is, with no electric flux) wrapped on associative
and special Lagrangian (SLAG) three-cycles. We recover the results for wrapped
NS5-branes in type IIA theory \cite{gmpw} 
in the special case that the vector constructed as a spinor bilinear is
Killing so that one can dimensionally reduce along this direction.
We also comment on the relationship of our approach with the
work of \cite{FS}. In particular, we give the supersymmetry constraints and
the non-linear PDEs (following from the Bianchi identity)
that one must solve to find solutions describing M5-branes
wrapped over associative and SLAG three-cycles. Furthermore, we discuss 
how our approach may be extended straightforwardly to obtain 
a similar description of five-branes wrapped on other calibrated cycles.

In section \ref{pureelectric} we discuss the case in which the internal
(magnetic) $G$-flux is swithed off. In this case our equations simplify 
drastically and we are able to give the most general solution.
In particular, we show that all AdS$_3$ solutions may be viewed as 
AdS$_4$ solutions, foliated by copies of AdS$_3$, with a weak 
$G_2$-holonomy manifold as internal space. We show how the compactifications
of \cite{Becker1,Becker2,AOG} are recovered in a degenerate limit in which the
internal spinor becomes chiral and, therefore, the $G_2$-structure becomes a 
$Spin(7)$-structure.

As illustration of our formalism in section \ref{dyonicsection} we give 
some explicit 
examples. We easily recover the dyonic M-brane solution
of \cite{ILPT}. This solution describes a $1/2$-BPS M5/M2 bound state and 
serves as a simple example of the essential features of our
geometries. We discuss also the relevance of our work to the 
recent ``dielectric flow'' solutions of \cite{warnerold,pope,warner}. These 
in fact also lie within our class of geometries. We present a class
of singular solutions based on $G_2$-holonomy manifolds, where the M5-brane
is completely smeared over the $G_2$-manifold.

Appendix \ref{g2appendix} gives a discussion of $G_2$-structures. Appendix
 \ref{m5} includes a 
brief discussion of the Hamiltonian formulation of the M5-brane theory. 
Appendix \ref{relations} contains
some relations useful in the main text.

\section{M-theory on eight-manifolds}

\label{oldstuff}

In this section we begin the analysis of eight-dimensional warped 
compactifications of M-theory. After summarising the {\it status quo} 
regarding the M2-brane-like solutions of \cite{Becker1, Becker2, AOG}, we 
then go on to describe how one extends
the analysis to allow for more general supersymmetric solutions with
fluxes.

The fields of eleven-dimensional supergravity consist
of a metric $\hat{g}_{MN}$, a three-form potential $C$ with field strength
$G =\diff C$, and a gravitino $\psi_M$. 
Supersymmetric backgrounds are those for which the  gravitino vanishes
and there is at least one solution to the equation
\bea
\delta \psi_M =   \hat{\nabla}_M \eta - \frac{1}{288} \left( G_{NPQR}
\hat{\Gamma}^{NPQR}{}_M- 8G_{MNPQ}\hat{\Gamma}^{NPQ} \right)\eta = 0~.
\label{basicsusy}
\eea
Here $\eta$ is a spinor of $Spin(1,10)$, and $\hat{\Gamma}_M$ form a 
representation of the eleven-dimensional Clifford algebra, 
$\{\hat{\Gamma}_M,\hat{\Gamma}_N\} = 2 \hat{g}_{MN}$.
We take the spacetime signature to be $(-,+,\ldots,+)$, so that one may take 
$\hat{\Gamma}_M$ to be hermitian for $M\neq0$ 
and anti-hermitian for $M=0$. Geometrically, \reef{basicsusy} is a parallel transport equation for a 
generalised connection, taking values in the full Clifford algebra, 
whose holonomy lies in $SL(32,\bbR)$ \cite{hull}. 
In our conventions the equations of motion are 
\bea
\hat{R}_{MN} -\frac{1}{12} (G_{MPQR}\hat{G}_{N}{}^{PQR}-\frac{1}{12}
\hat{g}_{MN}G_{PQRS}\hat{G}^{PQRS}) & = & 0~{}\label{einstein}\\
\diff \,\hat{*}\, G + \frac{1}{2} G \wed G & = & 0~.\label{G-equation}
\eea
One also has the Bianchi identity $\diff G=0 $. Generically the field
equations \reef{einstein} and \reef{G-equation} receive higher order 
corrections. In particular, the latter equation has a contribution 
$X_8$ on the right hand side, where
\bea
X_8  &=&  -\frac{(2\pi)^2}{192}\left(p_1^2-4p_2\right)~.
\eea
Here $p_i$ is the $i^{\mathrm{th}}$ Pontryagin form, and we have set 
the M2-brane tension equal to one. 


We will consider supersymmetric geometries with Poincar\'e or AdS invariance
in three external dimensions. Thus a general such ansatz for the metric is 
of the form
\bea
d \hat{s}^2_{11}  & = & \ex^{2\Delta} (ds^2_3+ g_{mn} dx^m dx^n)
\eea
and for the $G$-field we take the maximally symmetric ansatz 
\bea
G_{\mu\nu\rho m} & =&  \e_{\mu\nu\rho} g_m \nn
G_{mnpq} & & \mathrm{arbitrary} ~,
\eea
where here, and henceforth, Greek indices run over $0,1,2$ and Latin indices 
run over $3,\dots 10$ -- that is, over the internal manifold. 
We adopt the standard realisation of the eleven-dimensional Clifford algebra
Cliff$^{\mathrm{even}} (\bbR^{1,10})\simeq$   Mat$ (32,\bbR) \simeq $ Cliff $ 
(\bbR^{1,2})\otimes $Cliff $(\bbR^{0,8})$, namely
\bea
\hat{\Gamma}_\mu = \ex^{\Delta}(\gamma_\mu \otimes \gamma_9)\nn
\hat{\Gamma}_m = \ex^\Delta(\id \otimes \gamma_m)
\label{gammas}
\eea
A convenient explicit representation of the three-dimensional Clifford algebra
is given by $\gamma_0 = i\sigma_1, \gamma_1=\sigma_2, \gamma_2=\sigma_3$, 
where $\{\sigma_k\mid k=1,2,3\}$ are the Pauli 
matrices. The eight-dimensional gamma-matrices are $16\times 16$ real,
symmetric matrices. 
We have also $\gamma_9^2=\id$.
An eleven-dimensional spinor $\eta $ is likewise decomposed into three and 
eight-dimensional spinors as
\bea
\eta  & = & \psi \otimes \xi~.
\eea
The Majorana condition in eleven dimensions then imposes the following 
reality constraints:
\bea
\psi^* = \gamma_2 \psi~, \qquad \qquad {\xi}^* = \xi~.
\eea
Thus $\psi$ has two real components, and $\xi$ has {\em sixteen} real  
components.
The supersymmetry equation of interest \reef{basicsusy} may now be decomposed 
into two parts
\bea
\delta \psi_\mu  &=& \nabla_\mu \eta  +
\frac{1}{6}\ex^{-3\Delta}(\gamma_\mu \otimes g_m \gamma^m) \eta
- \frac{1}{2} (\gamma_\mu \otimes \de_m \Delta \gamma^{m}\gamma_9)\eta\nn
&& - \frac{1}{288} \ex^{-3\Delta}
(\gamma_\mu \otimes G_{npqr}\gamma^{npqr}\gamma_9) \eta\label{grav:ext}\,=\,0\\
\delta \psi_m  &=& \nabla_m \eta  + \frac{1}{2} (\id \otimes
\gamma_m{}^n\de_n \Delta)\eta + \frac{1}{12} \ex^{-3\Delta}(\id \otimes
\gamma_m{}^n g_n \gamma_9) \eta - \frac{1}{6} \ex^{-3\Delta} (\id \otimes
\gamma_9) g_m \eta\nn
&& -\frac{1}{288} \ex^{-3\Delta} \Big[ (\id \otimes G_{pqrs}
\gamma^{pqrs}{}_m)-8(\id \otimes G_{mpqr}
\gamma^{pqr}) \Big]\eta\,=\, 0
\label{grav:int}
\eea
which we refer to as the external and internal equations, respectively. 

In the rest of this section we will assume, as in  
\cite{Becker1,Becker2}, 
that the internal spinor is chiral. We will briefly review the consequences of
this restriction, before lifting it in the rest of the paper.
If $\xi$ is chiral, without loss of generality,
one may take $\gamma_9 \xi = \xi$. Requiring that 
$\nabla_\mu \psi = 0$ in \reef{grav:int} then implies
\bea
\left[-\frac{1}{288} \Delta_B^{3/2} G_{mpqr} \gamma^{mpqr} 
+\frac{1}{6} \Delta_B^{3/2} g_m \gamma^m 
+\frac{1}{4} \gamma^m \de_m \log \Delta_B
\right] \xi = 0 
\eea
where, for easier comparison with \cite{Becker1,Becker2}, 
we have defined the warp factor 
$\Delta= -\frac{1}{2}\log \Delta_B$. Projecting this equation onto its positive and negative chirality parts\footnote{Notice that this projection simplifies 
somewhat the analysis in the original papers \cite{Becker1,Becker2}.} we obtain
\bea
g_m = \de_m \Delta_B^{-3/2}~, \qquad \qquad G_{mpqr} \gamma^{mpqr} \xi  =  0~.
\label{project:singlet}
\eea
Upon rescaling the spinor and the internal metric as 
$\xi =  \Delta_B^{-1/4} \tilde{\xi}$ and 
$g_{mn}  =  \Delta_B^{3/2} \tilde{g}_{mn}$ respectively,
the relations (\ref{project:singlet}) allow one to simplify the internal 
part of the gravitino equation, yielding
\bea
\tilde{\nabla}_m \tilde{\xi} + \frac{1}{24}\Delta_B^{-3/4}G_{mpqr}
\tilde{\gamma}^{pqr}\tilde{\xi} & = & 0 ~.
\label{endofstory}
\eea
One again notes that the two terms in \reef{endofstory} have 
opposite chirality, and must therefore vanish separately. In particular it
follows that the metric $\tilde{g}_{mn}$ has $Spin(7)$ holonomy and the internal
flux satisfies
\bea
G_{mnpq} \gamma^{npq} \xi & = & 0~,
\label{spinor-27}
\eea
implying that some, but not all, 
of the $Spin(7)$ irreducible components of the flux must vanish. Recall that 
on manifolds with $Spin(7)$ structure four-forms may be decomposed  into 
four irreducible components \rep{70} $\to$ 
\rep{35} +  \rep{27}+ \rep{7} + \rep{1}  under $SO(8)\mapsto Spin(7)$ (see e.g.
\cite{salamon}). A convenient way to understand the condition
\reef{spinor-27} is to recast it into a tensorial equation \cite{AOG}.
Multiplying \reef{spinor-27} on  the left with $\xi^\trsp\gamma^r$ one
obtains 
\bea
T_{mn} \equiv\frac{1}{3!} G_{mpqr}\Psi^{pqr}{}_n & = & 0
\label{Tmn} 
\eea
where $\Psi$ is the Cayley four-form, characterising the $Spin(7)$ structure.
A general two-index tensor decomposes into the $SO(8)$ irreducible representations 
\rep{35} + \rep{28} + \rep{1}, which, under $SO(8)\mapsto Spin(7)$, further 
reduces to \rep{35} + \rep{21} + \rep{7} + \rep{1}. However, given the 
representation content of the four-form $G$, $T_{mn}$ 
must contain only the irreducible representations \rep{35} + \rep{7} + \rep{1}. 
One therefore concludes that only the 
\rep{27} component of the internal flux is allowed. A characterisation of this 
representation may also be given as follows
\bea
G_{\mathbf{27} \, mnpq} & = & \frac{3}{2}
G_{\mathbf{27}\, rs[mn}\Psi_{pq]}{}^{rs}~.
\eea

In conclusion, the general solution takes the form 
\bea
d \hat{s}^2_{11} & = &  H^{-2/3} \eta_{\mu\nu}dx^\mu dx^\nu +
 H^{1/3}\tilde{g}_{mn} dx^m dx^n\nn
G & = & \diff x^0 \wed \diff x^1 \wed \diff x^2 \wed \diff (H^{-1}) + G_{\mathbf{27}} 
\label{beckersolution1}
\eea
with the warp factor satisfying the equation
\bea
\tilde{\Box}H + \frac{1}{2} G_\mathbf{27} \wedge G_\mathbf{27} &
= & X_8
\label{BoxH}
\eea
where $G_{\mathbf{27}}$ is harmonic, and we have not included any explicit
space-filling M2-brane sources. Integrating \reef{BoxH} over a 
compact $X$ gives
\bea
\frac{1}{2}\int_X \frac{G_{\mathbf{27}}}{2\pi}\wedge \frac{G_{\mathbf{27}}}{2\pi} = 
-\frac{1}{192}\int_X p_1^2 - 4p_4  & = &  \frac{\chi(X)}{24}~.
\label{top}
\eea

In
general, existence of a nowhere vanishing section of a vector bundle requires 
that the Euler class of that bundle is zero.
Thus existence of a 
nowhere vanishing positive/negative chirality spinor requires that
$\chi(\bbS_{\pm})=0$, and it is this condition which 
gives the relation between
the topological invariants in the last equality in
\reef{top} (see, for example, \cite{LM}). One then has compact solutions 
with flux provided
the flux is quantised appropriately.

Note that these solutions describe M2-branes where the transverse space is 
a $Spin(7)$ holonomy manifold. 
Non-compact examples of such solutions may be found in \cite{Garyspapers} and
\cite{brito}. 
Notice that $G_{\mathbf{27}}$ decouples from the supersymmetry 
conditions, but it does play a role in the equations of motion, providing
the ``transgressive'' terms \cite{Garyspapers}.

The present analysis is readily extended to cases with more supersymmetry. For
example when   
$\xi$ is a {\em complex} chiral spinor \cite{Becker1} we have two 
$Spin(7)$ structures of the same chirality or, equivalently, an $SU(4)$-structure.
Repeating the same steps, one shows that the general solution is now of the 
form \reef{beckersolution1}, \reef{BoxH} with $\tilde{g}_{mn}$ having $SU(4)$
holonomy. The magnetic flux which drops out of the supersymmetry 
equations is given by $G_{(2,2)}$ (that is, the four-form has two holomorphic 
and two
anti-holomorphic indices with respect to the corresponding complex structure)
where $G_{(2,2)}$ is also primitive, so that taking the wedge-product with 
the K\"ahler form gives zero. 
Again, these solutions are akin to M2-branes transverse to Calabi-Yau 
four-folds,
and the role of the internal flux is to provide an additional source term in 
the equation for the warp-factor.

\subsection*{Generalisation}
\label{generalisation}

As we have summarised, imposing that the internal spinor 
be chiral leads to M2-brane-type solutions. However, there clearly should be
another way to obtain a supersymmetric Minkowski$_3$ vacuum from 
M-theory: one may wrap 
space-filling M5-branes over a supersymmetric three-cycle in a special 
holonomy manifold. Such cycles are calibrated. In particular, 
one may wrap the M5-branes over an associative three-cycle in a $G_2$-holonomy
 manifold (times a circle) to obtain an $\mathcal{N}=1$ vacuum, or
a special Lagrangian cycle in a Calabi-Yau three-fold (times a two-torus) to 
obtain an $\mathcal{N}=2$ vacuum. 
When one includes the
back-reaction of the brane on the initial geometry, one no longer has a 
manifold of special holonomy, but rather some more general geometry with
flux. However, the $G_2$ or $SU(3)$ structures still remain, respectively. 
Such manifolds admit two (respectively four) invariant Majorana-Weyl spinors, 
one (respectively two) of \emph{each} chirality. Thus to describe more general
supersymmetric solutions with fluxes one has to generalise the form of the
internal spinor. We will also find that when one lifts
the chirality assumption, one can find supersymmetric 
AdS$_3$ solutions, and we will present a simple class of examples 
in this paper.

From a more mathematical viewpoint, there is no reason to restrict the
spinor to be chiral. The M-theory Killing spinor equation is geometrically 
a parallel transport equation for a supercovariant connection 
taking values in the Clifford algebra Cliff$^{\mathrm{even}}
(\bbR^{1,10})\simeq$   Mat$ (32,\bbR)$. Indeed, in the three/eight split of the
eleven-dimensional spinor $\eta $, the internal spinor $\xi$ turns out to 
have 16 real components, {\it i.e.} it belongs to $Spin(8)_+\oplus Spin(8)_-$. 
We are therefore led to consider an internal 16-dimensional 
spinor of indefinite
chirality\footnote{For a four-seven decomposition, it was noticed in
\cite{deWit:1984nz}, and more recently also 
in \cite{ali,M7}, that in order to have non-trivial 
$G$-flux a generic spinor ansatz must be allowed.}, 
which in general can be written in 
the following form
\bea
\eta & = & \ex^{-\frac{\Delta}{2}}\psi \otimes (\xi_+ \oplus \xi_-) 
\eea
where $\gamma_9\xi_\pm=\pm\xi_\pm$ are real chiral spinors in eight dimensions,
 and $\psi$ is a Majorana spinor in three dimensions. The factor 
$\ex^{-\Delta/2}$ has been inserted for later convenience. 
For calculational convenience it is useful to introduce the
non-chiral 16-dimensional spinors
\bea
\e^+ & = & \frac{1}{\sqrt{2}} (\xi_+ + \xi_-)
\eea
and $\e^- \equiv \gamma_9 \e^+= (\xi_+ - \xi_-)/\sqrt{2}$. 
The advantage of working with $\e^\pm$, as opposed to
$\xi_\pm$, is that the former will turn out to have constant norms,
which, without loss of generality, we take to be unity, whereas 
the chiral spinors do not have this  
desirable property.

Since we wish to allow for AdS$_3$ compactifications in our analysis, we 
impose the following condition on the external spinor:
\bea
\nabla_\mu \psi + m \gamma_\mu \psi & = & 0~.
\eea
Writing the $G$-flux as 
\bea
G = \ex^{3\Delta} (F + \vol_3\wedge f)
\eea
with $F$ and $f$ parameterising the magnetic and electric components,
respectively, the supersymmetry conditions 
 may be written in terms of $\e^\pm$ as follows:
\bea
\nabla_m \e^\pm \pm \frac{1}{24} F_{mpqr}\gamma^{pqr}\e^\pm 
-\frac{1}{4}f_n\gamma^n{}_m \e^\mp \pm m \,\gamma_m\e^\mp & = & 0
\label{gravitino-like}\\
\frac{1}{2}  \de_m \Delta \gamma^m\e^\pm 
\mp \frac{1}{288} F_{mpqr} \gamma^{mpqr} \e^\pm 
-\frac{1}{6}f_m \gamma^m \e^\mp 
\mp m \,\e^\mp & = & 0~.
\label{dilatino-like}
\eea
These equations are the starting point for our analysis.

\section{Supersymmetry and the $G_2$-structure}
\label{bilinears}

In \cite{gmpw} it has been recognised that the notion of $G$-structures and their
intrinsic torsion provides a powerful technique for studying 
Killing spinor equations in the presence of fluxes. A 
rigorous 
account of the mathematics may be found, for example, in \cite{salamon}.
For our purposes, a $G$-structure in $d$ dimensions is a collection of
locally defined $G$-invariant objects, each in some irreducible representation
 of the (spin cover of the) tangent space group $Spin(d)\supset G$. Notice 
that, {\it a priori}, our equations need 
only be 
defined in some open set, which is why we use the term $G$-structure in this
local sense. When the objects in question extend globally over the whole 
manifold one has a $G$-structure in the stricter mathematical sense that 
the principal frame bundle admits a sub-bundle with fibre $G$. Of course,
there may be topological obstructions, and indeed the structure may break
down, for example at horizons.

The way that intrinsic torsion enters into the Killing spinor equations is 
via the fluxes. Exploiting this, one can study a supersymmetric geometry 
by extracting from the supersymmetry conditions the differential constraints 
on a set of forms that comprise the structure. These
forms may be constructed as spinorial bilinears. The intrinsic torsion
is an element of $\Lambda^1\otimes g^\perp$ (see, for example, \cite{intrinsic} or \cite{GP} for a brief review), which may be decomposed into irreducible
$G$-modultes, denoted $\mathcal{W}_i$ in this paper. The manifold will have
$G$-holonomy only when all the components vanish.

In the following we apply these methods to the case at hand, showing that 
one in general has a $G_2$-structure on the internal eight-manifold.
It is also important to establish what other conditions must be imposed 
on the structure for it to correspond to a solution of the supergravity
theory. We address this issue towards the end of the section.

We can construct explicitly a one-form, a three-form, and two four-forms
as bilinears in the spinors
\bea
\bar{K}_m & = &  \xi_+^\trsp \gamma_{m} \xi_- \nn
\bar{\phi}_{mnp} & = & \xi_+^\trsp \gamma_{mnp} \xi_-  \nn
\bar{\Psi}^\pm_{mnpr} & = & \xi_\pm^\trsp \gamma_{mnpr} \xi_\pm~.
\eea
In the calculations it is useful to re-express these in terms of the $\e^\pm$
spinors, and it is also useful to define the following auxiliary bilinear
\bea
Y_{mnpr}  =  \e^{\pm \trsp} \gamma_{mnpr} \e^\pm~.
\eea
Notice that, for a generic Clifford connection, 
the corresponding Killing spinors are not in general orthonormal, in contrast 
to the case of 
a connection on the $Spin (d)$ bundle \cite{intrinsic}. In particular, we
have that, using \reef{gravitino-like},
$\nabla(\e^{+\trsp}\e^+)=\nabla(\e^{-\trsp}\e^-)=0$. 
Thus we can normalise the spinors so as to obey
\bea
||\e^+||^2 = ||\e^-||^2 = \frac{1}{2}\left( ||\xi_+||^2 +||\xi_-||^2\right) =
1~.
\eea
On the other hand $\nabla(\e^{+\trsp}\e^-) \neq 0$, and we parameterise this 
non-trivial 
function, which takes values in the interval $[-1,1]$, as
\bea
\e^{+\trsp}\e^- & = &  \frac{1}{2}\left( ||\xi_+||^2 - ||\xi_-||^2\right) 
\equiv \sin \zeta ~.
\eea
It follows that the chiral spinors have norms 
$||\xi_\pm ||^2= 1\pm \sin \zeta$, and in the limit $\sin \zeta \to \pm 1$
one of the two vanishes.

 The stabiliser of each chiral 
spinor $\xi_\pm$ is $Spin(7)_\pm$, and their common subgroup 
is $G_2$.
In order to discuss the supersymmetry conditions in terms of the $G$-structure 
it is convenient to introduce rescaled forms,  defined as 
$\phi = (\cos\zeta)^{-1}\bar{\phi}$ and $K  =  (\cos\zeta)^{-1}\bar{K}$.
These are canonically normalised, namely 
 $||K||^2=1$, $||\phi ||^2=7$,  and define a $G_2\subset SO(8)$
structure in eight dimensions. 
One can give an explicit expression for $Y$ in terms of the other bilinears
\bea
Y & = & -i_K*\phi +  \phi\wedge K\sin\zeta~,
\eea
where here, and henceforth, $*$ denotes the Hodge dual on the internal 
eight-manifold.
The forms are also subject to the 
constraint 
\bea
i_K \phi & =&  0~.
\eea
Notice that $\phi$ defines a unique seven-dimensional metric via the equations
\bea
g^7_{ij} & = & (\det b)^{-\frac{1}{9}}\, b_{ij}~,\nn
b_{ij} & = & -\frac{1}{144} \epsilon^{m_1\dots m_7}
\phi_{im_1m_2}\phi_{jm_3m_4}\phi_{m_5m_6m_7}
\label{g2givesmetric}
\eea
where $\epsilon^{1234567}=1$, and we therefore have $g^{7}_{ij} K^j=0$.
The intrinsic torsion of the structure
lives in the space $\Lambda^1 \otimes g_2^\perp $
where $g_2 \oplus g_2^\perp= so(8)$. The Lie algebra $so(8)\simeq$ \rep{28}
decomposes as \rep{28} $\to  $2(\rep{7}) + \rep{14}, so the orthogonal
complement of the $g_2$ algebra is given by $g_2^\perp =$\rep{7} + \rep{7}. 
The intrinsic torsion then decomposes into ten modules  
\begin{equation}
\begin{aligned}
   T \in \Lambda^1 \otimes g_2^\perp 
      &= \bigoplus_{i=1}^{10} \mathcal{W}_i , \\
   (\mathbf{1} +  \mathbf{7} ) \times ( \mathbf{7} +\mathbf{7})  
      &\to 2(\mathbf{1})  + 4(\mathbf{7}) + 2(\mathbf{14})+ 2(\mathbf{27})~ .
\end{aligned}
\end{equation}
It turns out that the ten classes are determined by the exterior derivatives
of the forms. These have the following decompositions into irreducible $G_2$
representations 
\begin{equation}
\begin{aligned}
   \diff K 
   &\to  \mathbf{7}'' + \mathbf{7}''' +\mathbf{14}'\\
   \diff \phi  
      &\to \mathbf{1} + \mathbf{1}' +  \mathbf{7} +\mathbf{7}'  
       + \mathbf{27} + \mathbf{27}' \\
   \diff * \phi 
      &\to \mathbf{1}' + \mathbf{7} + \mathbf{7}' + \mathbf{14} + \mathbf{27}'~.
\end{aligned}
\end{equation}
Note that some representations appear more than once, and we have denoted
different representations with different numbers of primes. In particular, 
the representations
\rep{1} + \rep{7} + \rep{14} + \rep{27} are those relevant to $\diff_7 \phi$
and $\diff_7 *_7 \phi$ discussed in appendix \ref{g2appendix}. 
Using the identities (\ref{dphiproj1}) - (\ref{dphiproj3}) one shows
that $\de_K \phi$ and $\de_K *_7 \phi$ contain the same representations, denoted
with $\mathbf{1}' +\mathbf{7}' + \mathbf{27}' $. Finally, $\diff K = \alpha
\wedge K +
\beta$, with the one-form $\alpha$ corresponding to $\mathbf{7}'''$ and the 
two-form $\beta$ to $\mathbf{7}'' +\mathbf{14}'$. Notice that we  have an 
eight-manifold of $G_2$ holonomy if and only if 
$\diff K = \diff \phi = \diff *\phi=0$. Note also
that $K$ is Killing if and only if the representations
 $\mathbf{1}' +\mathbf{7}' + \mathbf{27}'
$ vanish. This follows on noticing that the non-trivial components of the Lie
derivative ${\cal L}_K g$ can be computed from 
${\cal L}_K \phi= i_K \diff \phi$  using equation \reef{g2givesmetric}.

We can proceed now to analyse the constraints imposed on the structure by the 
supersymmetry conditions. 
Rather than presenting all the details of the calculations, we shall instead 
present a simple illustrative computation. Consider, for instance,
$\nabla_r \bar{K}_m$. Using
 the definition of $\bar{K}$ as a spinor bilinear, together with 
the Killing spinor equations \reef{gravitino-like}, after some
straightforward gamma-algebra one calculates
\bea
\nabla_r \bar{K}_m = \frac{1}{12}F_{rijk} \e^{+\trsp} \gamma^{ijk}_{\ 
\ \ m} \e^+ - 2m \sin \zeta \, g_{rm} - \frac{1}{2}f^j \bar{\phi}_{jrm}~.
\label{covariantK}
\eea
Next, the first identity in appendix \ref{relations} (with the Clifford element
$A=\gamma_{rm}$), can be used to compute the antisymmetric part of 
\reef{covariantK}, 
obtaining  equation \reef{diff-K} below. Similar calculations yield the 
following constraints on the $G_2$ structure:
\bea
\diff (\ex^{3\Delta} K\cos\zeta )  & = & 0 \label{diff-K}\\
K \wed \diff (\ex^{6\Delta} i_K * \phi) & = & 0 \label{starphi}\\
\ex^{-12\Delta}\diff (\ex^{12\Delta}\vol_7 \cos\zeta) & = & -8m\, \vol_7 
\wedge K \sin\zeta \label{volume}\\
\diff \phi \wed \phi \,\cos\zeta\, & = & 24m\,\vol_7 
-4 *\diff \zeta+ 2\cos\zeta~ * f\label{dphiwedgephi}\label{dphivphi}
\eea
where $\vol_7=\frac{1}{7}\phi \wedge i_K * \phi$. 
The electric and magnetic components
of the flux are then determined as follows
\bea 
 \ex^{-3\Delta} \diff (\ex^{3\Delta}\sin\zeta) & = & 
 f -4mK\cos\zeta\label{electric}\\
\ex^{-6\Delta} \diff (\ex^{6\Delta}\phi\cos\zeta ) & = & - ~ * F  + 
F\sin\zeta
+ 4 m (i_K*\phi - \phi\wedge K\sin\zeta)\label{magnetic}~.
\eea
As we will discuss more extensively in section \ref{cal}, these 
equations
can be interpreted as generalised calibrations for membranes or fivebranes
wrapped on supersymmetric cycles (at least when $m=0$). An important point 
to emphasize is that
the conditions derived are also sufficient to ensure solutions to the Killing
spinor equations. Notice that generically $K$ {\em is not}
 a Killing vector. 
However, we see from \reef{diff-K} that it is in fact hypersurface orthogonal or,
equivalently, defines an integrable almost product structure \cite{intrinsic}   
which allows us to write the metric in the canonical form
\bea
d\hat{s}^2_{11} & = & \ex^{2\Delta (x,y)} (ds^2_3+ g^7_{ij}(x,y)dx^i dx^j)+ 
\frac{1}{\cos^2\zeta (x,y)}\ex^{-4\Delta(x,y)} d y^2~.
\eea
The remaining conditions may be thought of as putting constraints on the
seven-dimensional part of the $G_2$-structure. 
Consider, for example, equation \reef{starphi}.
From this we read off immediately that the \rep{14} representation is
absent and the \rep{7} is given by the Lee-form 
$W_4=18 \,\diff_7\Delta$. 
Likewise, equation \reef{volume} relates $\de_y \log\sqrt{g^{7}}$ to 
$\de_y \Delta$ and $\de_y\zeta$, hence fixing the $\mathbf{1}'$ representation.
 Continuing, 
the rest of the equations may be used to determine {\em all} the components
of the intrinsic torsion. One can thus construct a connection with non-trivial 
torsion which preserves the $G_2$ structure, and in particular preserves
two spinors of opposite chirality, corresponding to solutions of 
the supersymmetry 
equations. For simplicity we will present some details of the calculation
in the case of purely magnetic solutions in section \ref{puremagnetic}.

The four-form flux is completely determined in terms of the structure 
by \reef{electric} and \reef{magnetic}. In fact it is easy to show  that
there are no components which  automatically 
drop out of the supersymmetry equations 
\reef{gravitino-like} and \reef{dilatino-like}, in contrast to section 
\ref{oldstuff}. First let us decompose the four-form flux into 
$SO(7)$ irreducible representations:
\bea
F &= & F_4 + F_3 \wed K\label{F-SO7}~.
\eea
We thus want to check if there are $G_2$ irreducible components whose 
Clifford action $F_{mnpq}\gamma^{npq}$ annihilates both the spinors 
$\xi_\pm$, namely
$F_{4\, mnpq}\gamma^{npq} \xi_\pm  =
F_{3\,mnp}\gamma^{np} \xi_\pm  =  0$. This would imply
that the following tensors vanish
\bea
\frac{1}{2!}F_{3\,mpq}\phi^{pq}{}_{n} & = &  0\nn
\frac{1}{3!}F_{4\,mpqr} (*_7\phi)^{pqr}{}_{n} & = & 0~.
\eea
As discussed
in appendix \ref{g2appendix} these tensors contain all the components
of $F_3$ and $F_4$, which should therefore vanish identically. 
This situation is to be contrasted 
with the cases where we have spinor(s) of a fixed chirality, as recalled
in section \ref{oldstuff}.  
Each spinor defines a $Spin(7)$
structure and the \rep{27} component of the flux, with respect to that
structure, is undetermined by the supersymmetry equations. The existence of two
spinors with opposite chirality means that the associated 
$Spin(7)$ structures have opposite self-duality, and the undetermined flux 
should therefore simultaneously be in the \rep{27_+} and \rep{27_-}, 
and hence is trivial.

All the non-zero components of the flux can be extracted from the conditions 
\reef{diff-K} - \reef{magnetic}. As examples, and for later reference, let us
give the 
expressions for the \rep{1} and \rep{7} components of $F_3$ ({\it cf}. appendix
\ref{g2appendix})
\bea
\pi_1(F_3)  &=& \frac{2}{7}(\de_K\zeta -2m)\,\phi\nn   
\pi_7(F_3)  &=&-\frac{1}{2}\ex^{-3\Delta}\diff_7 
  (\ex^{3\Delta}\cos \zeta)\con i_K * \phi
\label{Hcomponents}
\eea
and of $F_4$
\bea
 \pi_1(F_4) &=& \frac{2}{7}\left(4m\sin\zeta - \ex^{-3\Delta}\de_K(\ex^{3\Delta} 
 \cos\zeta )\right) \,i_K * \phi\nn
 \pi_7(F_4) &= &\frac{1}{2}\phi \wedge \diff_7 \zeta~.
\label{F4components}
\eea

A solution will also have to obey the equations of motion and
Bianchi identity. Using the above expressions for the fluxes, it is 
straightforward to show that these reduce to the two  equations
\bea
\diff (\ex^{3\Delta} F) & = & 0\label{bianchiF}\\
\ex^{-6\Delta}\diff (\ex^{6\Delta}* f) + \frac{1}{2} F\wedge F & = & 0~.
\label{f-equat}
\eea
One can now show, using the results of \cite{GP},
 that the Einstein equation is automatically implied as an integrability
 condition for the supersymmetry conditions, once the $G$-field equation 
and Bianchi identity
 are imposed. It is useful to give explicitly 
 the external part of the Einstein
equation:
\bea
\ex^{-9\Delta}\Box_8 \ex^{9\Delta} - \frac{3}{2} ||F||^2 - 3 ||f||^2 + 72m^2 & 
= & 0~.
\label{exteinstein}
\eea
One may use this to prove that, when $m=0$, there are no compact solutions 
with electric and/or magnetic flux. 
Explicitly, one easily integrates \reef{exteinstein} over the compact manifold
$X$ to get 
\bea
\int_X \ex^{9\Delta}||F||^2 + 2\int_X \ex^{9\Delta}||f||^2  = 0
\label{no-go}
\eea
which requires $F=0$ and $f=0$. This is a rather general property of
supergravity theories \cite{dewit}.
The common lore to evade such ``no-go theorems'' is to appeal to higher 
derivative terms, such as the $X_8$ term mentioned in section \ref{oldstuff}, 
although these arguments typically neglect the corresponding terms 
in the Einstein equations. In our case, 
a non-zero $X_8$ seems to allow for the possibility of compact 
solutions\footnote{Equation \reef{no-go} receives a
correction proportional to $\int_X \ex^{3\Delta}\sin\zeta X_8$.}. One 
must then also satisfy 
\bea
\int_X G_\mathrm{internal} \wedge G_\mathrm{internal} & = & 0
\eea
which is implied by 
integrating equation \reef{f-equat}. Here one uses the fact that $X_8$ 
integrates to 
zero. This is so because the existence of two linearly independent spinors of
opposite chirality implies that $\chi(\bbS_{\pm})=0$. Equivalently, the
vector $K$ constructed from the spinors is nowhere vanishing, which 
implies that the Euler number of the eight-manifold is zero.

Comparing with the results reviewed in section \ref{oldstuff} we see that 
allowing the internal spinor to be non-chiral has a led to a substantially
enlarged number of possible geometries and fluxes. We emphasize the fact 
that AdS$_3$ solutions are not ruled out any more, and generically the 
internal manifold is not conformal to a $Spin(7)$ (or $SU(4)$) 
holonomy manifold. Note also that the function $\sin \zeta$ plays 
a role in our equations, and setting it to zero, or constant, 
rules out many supersymmetric geometries. In particular, from 
\reef{electric} it should be clear that $\sin \zeta$ is related to M2-brane 
charges, as we will see more explicitly in the next section.

\section{Generalised calibrations and dyonic M-branes}
\label{cal}

In this section we show how the supersymmetry constraints on the $G$-structure 
are related to a generalised calibration condition 
for the M5-brane. For simplicity we will restrict our analysis to Minkowksi$_3$
backgounds, and hence we set $m=0$ throughout this section.
We argue that the supersymmetric geometries we have been
describing so far may be thought of as being generated by
M5-branes wrapped over an associative three-cycle in a $G_2$-holonomy manifold.
An interesting twist to the story arises from the otherwise mysterious function
$\sin \zeta$, introduced in the last section.

Recall that the M-theory fivebrane has
a self-dual three-form field strength $H$ propagating on its world-volume, 
which 
induces an M2-brane charge on the M5-brane via a Wess-Zumino coupling. The
supergravity description of the M5-brane should account for this feature.
Thus we expect ``dyonic'' backgrounds -- that is, solutions 
with non-trivial electric {\em and}
magnetic fluxes. Placing  
a dyonic M-brane probe in its corresponding background should not then break
any further supersymmetry, and in particular a generalised calibration 
condition for such a probe should exist.
We will find that all of the supersymmetry equations 
(except for one) may be interpreted as generalised 
calibration conditions
for a probe M5-brane in our background. For example, \reef{magnetic} is
the generalisation of the associative calibration $\diff \phi=0$  in 
$G_2$-holonomy 
manifolds to dyonic M5-branes in warped backgrounds with flux.

Supersymmetric probes should saturate
a generalised calibration bound which minimises their energy.
In \cite{BLW} a calibration bound for the M5-brane was derived. 
Although some comments were made about general backgrounds the computation there
was for a flat space background with zero $G$-flux. It is easy to extend 
their analysis to the case of  non-zero $G$-flux, by taking into 
account  the Wess-Zumino terms. In appendix \ref{m5} we use the Hamiltonian 
formalism of \cite{townsend} to obtain an expression for the energy 
of a class of static M5-branes with non-zero background $G$-flux and 
world-volume
three-form $H$. This formula may then be used to show that 
supersymmetric branes are calibrated and saturate a bound on the energy.

The very alert reader may notice an obstacle in carrying out the above program.
The calibration bound derived in \cite{BLW} requires the existence of a
time-like Killing vector which in turn one uses to define the energy 
in a Hamiltonian formulation. Moreover, such a vector should arise 
as a spinor bilinear.
However, the supersymmetric geometries we are considering belong to the 
``null'' class, namely the stabiliser of the spinor $\eta$ 
(for any choice of $\psi$) is $\left(Spin(7)\ltimes \bbR^8\right)\times \bbR$
and the vector one constructs from it is a null vector \cite {M-wave,GP}. As
discussed in \cite{GP}, in this case the interpretation of the 
supersymmetry conditions as calibration conditions is less clear. 
However, by some sleight of hand, we may still use the static formulation of 
the M5-brane. The key to this is simply 
that we in fact have \emph{two} linearly independent
null spinors, from which we may construct a time-like Killing vector.

As discussed in appendix \ref{m5}, an M5-brane probe will be supersymmetric 
if, 
and only if,
\bea
\mathcal{P}_-\eta & = & 0
\eea
where $\mathcal{P}_-$ is a $\kappa$-symmetry projector, and $\eta$ is the
eleven-dimensional supersymmetry parameter.  
We have two linearly independent null spinors, $\eta_{\lambda} =
\sqrt{2}\ex^{-\Delta/2}\psi_{\lambda}\otimes \e^+$, where 
$\psi_{\lambda}$, for $\lambda=1,2$, are two linearly independent constant
spinors on $\bbR^{1,2}$. With an appropriate choice of $\psi_{\lambda}$, the 
vectors
one constructs from these spinors are $\partial/\partial t \pm 
\partial/\partial X_1$. Both vectors are null, but their sum  
$2k=2\partial/\partial t$ is time-like. 
Thus we are led to consider the following 
Bogomol'nyi-type bound:
\bea
\sum_{\lambda=1,2}||\mathcal{P}_-\eta_{\lambda}||^2  & = & 
\sum_{\lambda=1,2} \frac{1}{2}\eta_{\lambda}^{\dag}\mathcal{P}_-\eta_{\lambda}
\,\geq \,0~.
\eea
One then rewrites this bound in terms of the energy. 
From appendix \ref{m5} we have
\bea
E  & = &  T_{M_5}\left(\mathcal{C}_0 + \ex^{\Delta}L_{DBI}\right)
\label{energytext}
\eea
where $T_{M_5}$ is the M5-brane tension, $\mathcal{C}_0$ is the contribution
of a Wess-Zumino-like term to the energy, 
and $L_{DBI}$ is a Dirac-Born-Infeld action ({\it cf.} appendix \ref{m5}).
The bound may therefore be written
\bea
\ex^{\Delta}L_{DBI} \vol_5  & \geq & \sum_{\lambda=1,2} j^*\nu_{\lambda} + 
j^*\chi_{\lambda}\wedge H
\eea
where we have defined the space-time forms
\bea
\nu_{\lambda} & = & \frac{1}{(||\eta_1||^2+||\eta_2||^2)} \frac{1}{5!} 
\eta_{\lambda}^{\dag} \hat{\Gamma}_{0\ M_1 
\ldots M_5} \eta_{\lambda}  \diff X^{M_1}\wedge \ldots \wedge \diff X^{M_5}\nn
\chi_{\lambda} & = & -\frac{1}{(||\eta_1||^2+||\eta_2||^2)} \frac{1}{2!} 
\eta_{\lambda}^{\dag} \hat{\Gamma}_{0\ 
MN}\eta_{\lambda} \diff X^M\wedge \diff X^N
\label{su5}
\eea
and $j^*$ denotes a pull-back to the M5-brane world-volume. Using 
\reef{energytext} we obtain a bound on the energy density 
$\mathcal{E}= E \vol_5$:
\bea
\frac{1}{T_{M_5}} \mathcal{E}  & \geq & \sum_{\lambda=1,2} \left(j^*\nu_{\lambda} 
+ j^*\chi_{\lambda}\wedge H\right)
 + \mathcal{C}_0 \vol_5\label{bound}
\eea
where 
\bea
\mathcal{C}_0 \vol_5  & = & i_k C_6 - \frac{1}{2} i_kC\wedge (C-2H)
\eea
and a pull-back is understood on the right-hand side of this equation.

Given a static supersymmetric background, a pair $(\Sigma_5, H)$, with 
$\Sigma_5$ a 5-cycle and $H=h+j^* C$ a three-form 
on $\Sigma_5$ satisfying $\diff H = j^*G$, is said to be \emph{calibrated} if
the bound \reef{bound} is saturated on all tangent planes of $\Sigma_5$.
As we will show below, such a calibrated M5-brane worldspace then has minimal
 energy in its 
equivalence class $[(\Sigma_5, H)]$. Here, a pair $(\Sigma_5^{\prime},
H^{\prime})$ is in the same equivalence class as $(\Sigma_5, H)$ if 
$\Sigma_5$ is homologous 
to $\Sigma_5^{\prime}$ via a six-chain
$B_6$ (that is, $\partial B_6 = \Sigma_5 - \Sigma_5^{\prime}$) over 
which $H$ and $H^{\prime}$ extend to the same three-form, $H$, satisfying
$\diff H=j^*G$ on $B_6$. In fact, since $C$ clearly extends (it is defined over
all of space-time), it is enough to extend $h$ over $B_6$ as a closed form. 
Now, by Poincar\'{e} duality on the M5-brane worldvolume, $h$ defines a 
two-cycle $\Sigma_2 \subset \Sigma_5$, where $[\Sigma_2]$ is isomorphic
to $[h]$ under Poincar\'{e} duality. $h$ induces an M2-brane charge 
via the Wess-Zumino coupling \reef{WZ}, and thus $\Sigma_2$ may be thought
of as the effective M2-brane worldspace, sitting inside the M5-brane.

To prove the calibration bound on the energy the forms 
$\chi_\lambda$, $\nu_\lambda$ must obey suitable differential conditions. 
As we show
below, these combine to give the general conditions on the forms defining the 
$\left(Spin(7)\ltimes \bbR^8\right)\times \bbR$  structures 
in eleven dimensions \cite{GP}. These read\footnote{Our conventions
differ from those of \cite{GP}. To rectify this, one can simply change the 
sign of the gamma matrices of \cite{GP}. This leads to some extra
minus signs when using their results.} 
\bea
\diff \chi_{\lambda} & = & i_{\omega_{\lambda}} G\nn
\diff \nu_{\lambda} & = & i_{\omega_{\lambda}}\hat{*}G - \chi_{\lambda} 
\wedge G\label{su5eqns}
\eea
where in our case the one-forms
\bea
\omega_{\lambda} = \frac{1}{(||\eta_1||^2+||\eta_2||^2)}
\eta_{\lambda}^{\dag} \hat{\Gamma}_{0\ 
M}\eta_{\lambda} \ \diff X^M
\eea
are both null.
With our choice of $\psi_{\lambda}$, we may take their sum 
$\omega_1+\omega_2=-\diff t\ex^{2\Delta}$. The dual vector is then
simply $(\omega_1+\omega_2)^{\#} = \partial/\partial t =k$. A calibrated pair 
$(\Sigma_5,H)$ therefore obeys  
\bea
E(\Sigma_5, H) & = & \int_{\Sigma_5} \sum_{\lambda=1,2}\left(\nu_{\lambda} 
+\chi_{\lambda}\wedge H\right) +i_kC_6
-\frac{1}{2}i_kC\wedge (C-2H)\nn
& = & \int_{B_6} \sum_{\lambda=1,2}\left(\diff \nu_{\lambda} + \diff 
(\chi_{\lambda}\wedge H)\right) + \diff (i_kC_6)
- \frac{1}{2}\diff\left(i_kC \wedge(C-2H)\right)\nn
& & + \int_{\Sigma_5^{\prime}} \sum_{\lambda=1,2}\left(\nu_{\lambda} +
\chi_{\lambda}\wedge H^{\prime}\right) +i_kC_6
-\frac{1}{2}i_kC\wedge (C-2H^{\prime})\nn
& = & 0 +  \int_{\Sigma_5^{\prime}} \sum_{\lambda=1,2}\left(\nu_{\lambda} +
\chi_{\lambda}\wedge H^{\prime}\right) +i_kC_6
-\frac{1}{2}i_kC\wedge (C-2H^{\prime})\nn
& \leq & E(\Sigma_5^{\prime}, H^{\prime})
\eea
for any $(\Sigma_5',H')$ in the same equivalence class as $(\Sigma_5,H)$.
Notice that we have used, for example, $\diff (i_kC_6) = -i_k(\diff C_6) = -i_k
\left(\hat{*}G + \frac{1}{2}C\wedge G\right)$, in order to show that the
integral over $B_6$ vanishes.

Note also that this result holds for all cases where it is possible to 
construct 
an appropriate time-like Killing vector from the Killing spinors (not
necessarily as a bilinear), and thus it holds in particular for the entire
``time-like'' class of \cite{GP}.

It is now a simple matter to relate this to the supersymmetry equations 
of the last section. Indeed, these are equivalent to 
 \reef{su5eqns} on rewriting them 
in terms of the quantities defined in the last section. In particular, we have
that
\bea
\nu_1+\nu_2 & = & -\vol_2\wedge \ex^{6\Delta}\phi\cos\zeta -\diff t\wedge 
\ex^{6\Delta}Y\\
\chi_1+\chi_2 & = & +\vol_2 \ex^{3\Delta}\sin \zeta +\diff t \wedge 
\ex^{3\Delta} K \cos \zeta
\eea
where $\vol_2 = \diff X^1\wedge \diff X^2$ is the spatial two-volume.
Thus we have 
\bea
\diff (\chi_1+\chi_2) = \vol_2 \wedge 
\diff\left(\ex^{3\Delta}\sin\zeta\right) 
 -\diff t \wedge \diff\left(\ex^{3\Delta} K \cos \zeta\right) = i_kG = 
\vol_2 \wedge \ex^{3\Delta} f
\eea
which shows the equivalence of \reef{diff-K} and \reef{electric} with 
the first equation in \reef{su5eqns}, and also 
\bea
\diff (\nu_1+\nu_2) & = & -\vol_2 \wedge \diff \left(\ex^{6\Delta}\phi\cos\zeta
\right) + \diff t\wedge \diff\left(\ex^{6\Delta}Y\right)\nn
& = & i_k\hat{*}G -(\chi_1+\chi_2)\wedge G \nn
& = & -\vol_2\wedge\left(-\ex^{6\Delta}*F + 
\ex^{6\Delta}\sin \zeta F\right)-\diff t\wedge \ex^{6\Delta}\cos\zeta F\wedge K~.
\eea
This equation is clearly equivalent to the condition \reef{magnetic} together
with 
\bea
\ex^{-6\Delta}\diff\left(\ex^{6\Delta}Y\right) & = & -F\wedge K\,\cos\zeta~.
\label{dY-eq}
\eea
On expanding the various terms, this can be shown to be equivalent to 
\reef{starphi}, \reef{volume}, and the contraction of \reef{dphivphi} with $K$.
The relation \reef{nice} is useful for establishing this result.

Interestingly, \reef{electric} and \reef{diff-K} may also be 
derived from considerations of the M2-brane. 
In fact \cite{GP}, the first condition in \reef{su5eqns} is a generalised 
calibration condition for the M2-brane world-volume theory. The 
latter is more straightforward than the M5-brane theory as 
there is no form-field propagating on the M2-brane. Specifically,
there is a simple Nambu-Goto term plus the Wess-Zumino
electric coupling to the $C$-field. In this case, the energy is essentially
just the action. Equation \reef{electric} is then a 
calibration condition for a space-filling M2-brane, whereas \reef{diff-K}
is a calibration condition for an M2-brane wrapped over the $K$-direction. 
Notice that the remaining component of equation \reef{dphivphi} did not enter
the M5-brane calibration and in fact its eleven dimensional origin is in
the equation (2.18) of \cite{GP} for the Killing one-form $\diff k$. We 
suspect that
this should ultimately be related to a ``calibration'' for momentum carrying
branes, or waves. It would be interesting to understand this point further.

\section{M5 branes wrapped on associative and SLAG three-cycles}
\label{puremagnetic}

In this section we specialise our results to the case in which the electric 
component of the flux $f$ is set to zero as well as the mass $m$. This
situation corresponds to purely magnetic M5-branes wrapping three-cycles 
inside the transverse eight-manifold, with vanishing 
world-volume three-form field $H$. The geometries we consider are 
then of the form $\bbR^{1,2}\times M_8$, where $M_8$ generically admits a 
$G_2$ structure corresponding to ${\cal N}=1$ in the external Minkowski$_3$
space, or an $SU(3)$ structure corresponding to ${\cal N}=2$. 
We will also briefly discuss how one can easily extend these results to the 
case of M5-branes wrapping various four-cycles.

\subsection*{Associative calibration and ${\cal N}=1$}

Specialising the equations of section \ref{bilinears} to the case at hand
we get the following set of conditions on the $G_2$ structure:
\bea
\diff (\ex^{3\Delta} K)  & = & 0\label{ns1}\\
K \wed \diff (\ex^{6\Delta} i_K *\phi) & = & 0\label{ns3}\\
\diff (\ex^{12\Delta}\vol_7) & = & 0\label{newns}\\
\diff \phi \wed \phi & = &
0\label{ns2}\\
\ex^{-6\Delta} \diff (\ex^{6\Delta}\phi) & = & 
- ~ * F  ~.\label{m0:flux}
\eea
The metric takes the following form 
\bea
d \hat{s}^2_{11} & = & \ex^{2\Delta} \left(ds^2 (\bbR^{1,2}) + ds^2_7\right) +
\ex^{-4 \Delta } dy^2~.
\eea
Notice that equation \reef{newns} is equivalent to
$ \de_y \log\sqrt{g^7} =  -12 \, \de_y \Delta$. Thus M5-branes
wrapped on associative three-cycles give rise to an almost product structure 
geometry on the transverse eight-manifold which, at any fixed value of $y$,
admits a $G_2$ structure of the type ${\cal W}_3\oplus {\cal W}_4$. 
Explicit solutions 
were presented in \cite{AGK}. The close relation
to the results of \cite{gmpw} is of course not accidental.  
Recall that $K$ is generically not a Killing vector. However, when it is, 
one can Kaluza-Klein reduce along the $y$ direction (identifying the dilaton as
$\Phi=-3\Delta$) to get solutions of the type IIA theory,
which describe NS5-branes wrapped on associative three-cycles \cite{gmpw}. 
Of course, if additional Killing vectors are present in 
specific
solutions one can also reduce along those directions to obtain 
type II backgrounds which may contain RR fluxes in addition to the NS 
three-form. 

Let us comment here on the relationship of our approach to 
the work initiated in \cite{FS} and expanded upon in a series of papers (see
\cite{mrsmith} for a review). The strategy
in \cite{FS} is to write down an appropriate ansatz for the solution and then 
substitute this into the supersymmetry equations. Eventually 
one is left with a non-linear PDE for some metric functions which parameterise 
the ansatz (after imposing the Bianchi identity). It should be
clear that using the techniques of $G$-structures one can easily 
recover the 
various constraints obtained using the approach of \cite{FS}. As a
bonus we have in addition a physical interpretation of 
the constraints in terms of generalised 
calibrations\footnote{The relation of the work of \cite{FS}
to generalised calibrations was noticed in
 \cite{cio,husain1,husain2,castoro} . These papers consider
a class of geometries where the internal space is Hermitian. This is related 
to the fact that these geometries describe M5 or M2 branes wrapped on
holomorphic cycles.}
and, thanks to the
machinery of intrinsic torsion, we can apply the technique to more 
general cases which 
do not admit complex geometries. The work of \cite{ALE}, using the 
$G$-structure approach, recovers
the ${\cal N}=1$  geometries of \cite{FSN=1}, corresponding to M5-branes
wrapped on K\"ahler two-cycles in Calabi-Yau three-folds (times $S^1$), 
{\it i.e.} seven-manifolds with $SU(3)$-structure, after including the flux
back-reaction.
These in turn reduce in type IIA to the complex geometries first described in 
\cite{hulltorsion,strominger} in the context of Type I/Heterotic, 
as can easily be checked using the equivalent formulation given in
\cite{intrinsic}. It is straightforward to see that a similar formulation
exists for the ${\cal N}=2$ geometry of \cite{FS} corresponding to M5-branes 
wrapped on  K\"ahler two-cycles in seven-manifolds with $SU(2)$-structure.
In this case the supersymmetry conditions are exactly those discussed in the 
type IIA limit in section 6 of \cite{intrinsic}, with the transverse space
$\bbR^2$ replaced by $\bbR^3$. Clearly, all the geometries  
discussed in \cite{intrinsic} have a direct counterpart in M-theory as 
wrapped M5-branes.

Thus, imposing the Bianchi identity on $G$, we can write down
the associative analogue of the non-linear equations of \cite{FS}, which reads
\bea
\diff_7 \left[ \ex^{-6\Delta} *_7 \diff_7 (\ex^{6\Delta}\phi)\right]+ \de_y^2
(\ex^{6\Delta} *_7 \phi) & = & 0 
\label{assPDE}
\eea
where we have used the following expression for the $G$ field
\bea
G & = & \de_y (\ex^{6\Delta} *_7 \phi) + \ex^{-6\Delta} *_7 \diff_7
(\ex^{6\Delta}\phi)\wed \diff y~. 
\eea
This is equivalent to the generalised calibration condition \reef{m0:flux}. 
Here we do not write down possible source terms.
Note that equation \reef{G-equation} is automatically satisfied, with
$G\wed G$  and $\diff \,\hat{*}\, G$ being separately zero (using \reef{ns2},
\reef{m0:flux}, respectively).

Next, as promised in section \ref{bilinears}, we address more explicitly 
the issue of sufficiency of the conditions we have derived. This is ensured 
by the
careful counting of irreducible components of the intrinsic torsion, but it 
is perhaps instructive to look also at the Killing spinor equations directly.
The strategy is essentially to substitute our conditions back into the Killing
 spinor equations and check that they indeed admit solutions. 
Substituting the conditions \reef{ns1} - \reef{m0:flux} into the supersymmetry  
equations, we find that the external part \reef{dilatino-like} gives
\bea
- 3 \gamma^i\de_i \Delta \, \xi_\mp + \frac{1}{12} F_{3\,ijk} \gamma^{ijk} \xi_\mp +
\frac{1}{48} F_{4\; ijkl} \gamma^{ijkl}\xi_\pm - 3\ex^{3\Delta}\de_y \Delta
\xi_\pm & = & 0
\label{Kill:dilatino}
\eea
while the internal  part \reef{gravitino-like} gives 
\bea
\nabla_i^{(7)} \xi_\pm + \frac{1}{8} F_{3\,ijk} \gamma^{jk} \xi_\pm+ \frac{1}{4} 
\ex^{3\Delta}\de_y ( g^7_{ij} ) \gamma^j \xi_\mp + \frac{1}{24} F_{4\,
ijkl}\gamma^{jkl} \xi_\mp & = & 0\label{Kill:a-comp}\\
  \de_y \xi_\pm +
\frac{1}{4} e^i{}_{[a}\de_y\, e_{b]i} \gamma^{ab} \xi_\pm & = &
0\label{Kill:y-comp}
\eea
where here the indices run from $1$ to $7$ and $\nabla^{(7)}$ is the 
Levi-Civita
connection constructed from $g_{ij}^7$. Next, we can simplify these equations
using the fact that 
\bea
\frac{1}{3!}F_{4\,iklm} (*_7\phi)^{klm}{}_{j} & = & - \ex^{3\Delta} \de_y
 ( g_{ij}^7) 
 \label{F4=1+27}
\eea
which can be  computed from the expression for the flux
\reef{m0:flux} and the conditions  \reef{newns}, \reef{ns2}.
Notice that, as discussed in appendix \ref{g2appendix}, 
 this means that the \rep{7}
representation in $F_4$ vanishes, as is implied by the second equation in
\reef{F4components}. One can then show that the equations 
\reef{Kill:dilatino} and \reef{Kill:a-comp} reduce respectively to 
\bea
 \gamma^i\de_i \Phi \, \xi  + \frac{1}{12} F_{3\,ijk} \gamma^{ijk} \xi
& = & 0\nn
\nabla_i^{(7)} \xi + \frac{1}{8} F_{3\,ijk} \gamma^{jk}\xi & = & 0
\eea
where $\xi$ is the unique seven-dimensional spinor corresponding to $\xi_\pm$
in eight dimensions, and we have intentionally used the notation 
$\Phi=-3\Delta$ to demonstrate
that the resulting equations are essentially the dilatino and
gravitino equations of type IIA. Thus, by the results of \cite{f+i,ivanov1,gmpw}, 
we indeed have a
solution. Equation \reef{Kill:y-comp} is solved by taking the spinor to be
$y$-independent and the $\omega_{yab}$ component of the spin-connection 
to be in the \rep{14} of $G_2$: this simply corresponds to the standard 
 choice of local frame where $\phi_{abc}$ has constant coefficients.

\subsection*{SLAG calibration and ${\cal N}=2$}

Following the same line of reasoning as above, the equations describing 
M5-branes wrapping SLAG three-cycles in manifolds with an 
$SU(3)$-structure may almost be extrapolated from those pertaining to
NS5-branes 
wrapping the same cycles obtained in \cite{gmpw}. By repeating 
the arguments of \cite{gmpw,intrinsic} we have that doubling the amount of
supersymmetry yields the presence of two $G_2$ structures, whose maximal common
subgroup gives us an $SU(3)$-structure. One may then 
carry over the previous analysis by considering a Killing spinor
of the type $\psi \otimes (\xi_+ \oplus \xi_-)$
where $\psi$ and $\xi_\pm$ are now {\em complex} spinors. Thus one 
can also  
think of $SU(3)$ as arising from two $SU(4)$-structures having 
opposite chiralities, each defined by a complex Weyl spinor. 
Notice that this geometry then belongs to both the ``null''
and ``time-like'' classes of \cite{GP}, as $SU(3)$ embeds into 
$(Spin(7)\ltimes \bbR^8)\times \bbR$ as well as into $SU(5)$.

In a real notation, we take our spinors to be
\bea
\eta^{(a)} & = & \ex^{-\frac{\Delta}{2}}\psi^{(a)} \otimes 
(\xi_+^{(a)} \oplus \xi_-^{(a)}) \qquad \quad a=1,2
\eea
where each of the $\psi^{(a)}$ has two independent real components, thus 
corresponding to ${\cal N}=2$ in three dimensions. To realize 
the $SU(3)$ structure explicitly one can now construct
additional bilinears. We refer to appendix B of \cite{intrinsic}
for details. Notice that we have two vectors, which in a local frame
are given by $K^{(1)}=e^7$, $K^{(2)}=e^8$ and a two-form given by
\bea
J_{mn} & = & \e^{+\trsp}_{(1)}\gamma_{mn}\e^{+}_{(2)},
\eea
where, as before, $\e^{+}_{(a)}= (\xi_+^{(a)} + \xi_-^{(a)})/\sqrt{2}$ and
in a local frame we have $J = e^{12}+e^{34}+e^{56}$.
There are, of course, other bilinears that 
one can consider, but this is all we need. In fact, in terms of the associative
three-forms, we have 
\bea
\phi^{(a)} & = &  J \wedge K^{(1)} \pm  \mathrm{Im} \, \Omega
\eea
with $\Omega = (e^1+ie^2)\wedge (e^3+ie^4)\wedge (e^5+ie^6)$.
The $SU(3)$ structure is given by $K^{(a)}$, $J$, $\Omega$ with the last two
defining the structure in its canonical dimension of six, and $i_{K^{(a)}} J=
i_{K^{(a)}}\Omega =0$.  
 
Using the Killing spinor equations, after
some calculations one arrives at 
the following set of conditions:
\bea
\diff (\ex^{3\Delta} K^{(a)})  & = & 0\label{su3:1}\\
\diff (\ex^{3\Delta} J) & = & 0\label{su3:2}\\
K^{(1)}\wedge K^{(2)} \wedge \diff (\ex^{3\Delta}\mathrm{Re}\, \Omega) & = & 0
\label{su3:3}\\
\diff  (\mathrm{Im}\, \Omega)\wed \mathrm{Im}\, \Omega  & = &
0\label{su3:4}\\
\ex^{-6\Delta} \diff (\ex^{6\Delta} \mathrm{Im}\, \Omega ) & = & 
- ~ * F  \label{su3:flux}~.
\eea
The two vectors give rise to an almost product metric structure of the form 
\bea
d \hat{s}^2_{11} & = & \ex^{2\Delta} \left(ds^2 (\bbR^{1,2}) + ds^2_6\right) +
\ex^{-4 \Delta } (dy^2+ dz^2)~.
\eea
As discussed in \cite{intrinsic} the six-dimensional slices at fixed $y$
and $z$ have an $SU(3)$ structure with intrinsic torsion lying in the class
${\cal W}_2\oplus {\cal W}_4 \oplus {\cal W}_5$ with warp-factor 
$6\,\diff_6 \Delta = - W_4 = W_5$ (see \cite{chiossi,intrinsic} for details about 
the intrinsic torsion of $SU(3)$ structures). Notice that 
these geometries are {\em not}
Hermitian, which mirrors the fact that the M5-branes wrap SLAG three-cyles: 
equation \reef{su3:flux} is the corresponding generalised calibration
condition. 
Explicit solutions of this type were presented in \cite{GKW}.
The proof that the above equations are also sufficient to ensure the existence
of four solutions to 
the Killing spinor equations amounts to the observation that with these one 
can construct two $G_2$ structures, as in the previous subsection, each of 
which corresponds to two Killing spinors with opposite chiralities.

As in the previous case, let us write down the equation implied by the 
Bianchi identity $\diff G=0$. This is the SLAG-3 analogue of the 
equations of \cite{FS} and reads
\bea
\diff_6 \left[ \ex^{-9\Delta} *_6 \diff_6 (\ex^{6\Delta}\mathrm{Im}\,\Omega)
\right]+ \triangle_{yz}(\ex^{3\Delta} \mathrm{Re}\,\Omega) & = & 0 ~.
\eea
Where $\triangle_{yz}=\de_y^2+\de_z^2$ is the flat Laplacian in the transverse
directions. To derive this equation we have made use of the conditions above
to rewrite the flux in the following form
\bea
G  =  - \ex^{-9\Delta} *_6 \diff_6 (\ex^{6\Delta}\mathrm{Im}\,\Omega) \wedge
\diff y \wedge \diff z + \de_z (\ex^{3\Delta} \mathrm{Re}\, \Omega) \wedge \diff
y - \de_y (\ex^{3\Delta} \mathrm{Re}\, \Omega) \wedge \diff z~. 
\eea
The $G$ equation of motion \reef{G-equation} is again automatically satisfied.

\subsection*{More wrapped M5-branes}

We have presented the general conditions on the geometry 
of M5-branes wrapped on associative and SLAG three-cycles, giving 
explicitly the non-linear PDE which results from imposing the Bianchi identity.
M5-branes wrapped on K\"ahler two-cycles in Calabi-Yau two-folds and
three-folds were described in \cite{FS,FSN=1}, and in \cite{ALE} 
from the point of view of $G$-structures. Consulting the tables in
\cite{intrinsic} one realises that to complete the analysis of wrapped 
M5-branes one needs to consider four-cycles, yielding geometries of the type
$\bbR^{1,1}\times M_9$. Clearly, it is straightforward 
to extend our analysis to cover all the remaining cases of 
M5-brane configurations wrapping supersymmetric cycles. 
These will essentially be the M-theory lifts of the
conditions derived in \cite{intrinsic} for all possible wrapped NS5-branes in
the type IIA theory. For instance, we anticipate that, for static 
purely magnetic M5-branes, the flux is given by the generalised calibration condition
\bea
~ *_9 F & = & \ex^{-6\Delta} \diff (\ex^{6\Delta} \Xi) 
\eea
where $\Xi$ is the relevant calibrating form. Thus when fivebranes wrap
coassociative four-cycles in $G_2$-manifolds (times $\bbT^2$)  we have 
$\Xi=*_7\phi$; for K\"ahler four-cycles $\Xi=\frac{1}{2}J \wedge J$, 
and so on. Imposing the Bianchi identity gives the corresponding
non-linear PDE. Notice that the ``time-like'' case in \cite{GP} covers 
the case of M5-branes wrapped on SLAG five-cycles in Calabi--Yau five-folds,
and the resulting $SU(5)$  structure is described there in detail.

\section{All purely electric solutions}
\label{pureelectric}

In this section we discuss supersymmetric solutions with no internal 
components of the flux; namely, we set $F=0$. Suppose first that $m\neq 0$. In 
this case, setting to zero the \rep{1} and \rep{7} components of the flux in
\reef{Hcomponents} and \reef{F4components}  one can solve for $K$, $f$ 
and $\Delta $ in terms of the
 function $\zeta$, which one may take as a coordinate on the internal space, 
thus obtaining
\bea
K & = & \frac{1}{2m} \diff \zeta\nn
f & = & 3 \sec\zeta \diff \zeta\nn
\ex^{-\Delta} & = & \cos\zeta~.
\eea
Using these, one finds that the 
supersymmetry conditions \reef{diff-K} - \reef{magnetic} 
reduce to the single equation 
\bea
\diff (\ex^{3\Delta}\phi) & = & 4m\, \ex^{4\Delta} i_K * \phi~.
\label{confweak}
\eea
We can now define a conformally rescaled three-form 
$\tilde\phi = \ex^{-3\Delta}\phi$, and the corresponding 
four-form and metric 
$ \tilde *_7\tilde\phi = \ex^{-4\Delta} *_7\phi$ and $\tilde{g}_{mn} =
\ex^{-2\Delta}g_{mn}$, in terms of which equation \reef{confweak} becomes
\bea
\diff \tilde\phi & = & 4m\, \tilde{*}_7 \tilde \phi~.
\label{weakg2}
\eea
The genereral solution is therefore given by
\bea
d \hat{s}^2_{11} & = & \sec^2\zeta \left(ds^2_3 (\mathrm{AdS}_3)
+\frac{1}{4m^2}d\zeta^2\right) + d\tilde{s}^2_7\nn
G & = & 3 \vol_3 \wedge \sec^4 \zeta \diff \zeta\label{somethingelse}
\eea
where the seven-dimensional metric has weak $G_2$ holonomy, as dictated by 
\reef{weakg2}.
Notice that the $G$ equation of motion \reef{f-equat} is automatically 
satisfied since $\ex^{6\Delta}* f = 6m \,\tilde{\vol}_7$.

Compactifications of M-theory on weak $G_2$ manifolds 
were studied extensively in the 1980's (see, for example, \cite{DNP}). The 
simplest example is the well-known AdS$_4\times S^7$  compactification, which is
in fact maximally supersymmetric. Indeed, by a suitable change of
coordinates, one can check that 
the solution \reef{somethingelse} is of the form AdS$_4\times M_7$, where $M_7$ has
weak $G_2$ holonomy. Setting
$\sec\zeta =  \cosh (2mr) $,
the eleven-dimensional metric becomes
\bea
d \hat{s}^2_{11} & = & \cosh^2 (2mr) \,  ds^2(\mathrm{AdS}_3)
+d r^2  + d\tilde{s}^2_7~.
\eea
The four dimensional piece is the metric on AdS$_4$ with radius $l=1/2m$, 
foliated with copies of AdS$_3$. The seven-metric $d\tilde{s}^2_7$ is a weak
$G_2$ manifold, with metric normalised such that the Ricci tensor satisifies 
$\mathrm{Ric}=6m^2\tilde{g}$.

Let us consider briefly the case when $m=0$, so that the three-dimensional
external space is flat $\bbR^{1,2}$. In this case, setting to zero the
components of the internal flux \reef{Hcomponents} and \reef{F4components} 
implies that $\sin\zeta = \pm 1$. This is the limit in which one of the 
chiral spinors vanishes, leaving only the spinor of opposite chirality. The 
one-form $K$ and the three-form $\phi$ are then identically zero, while 
there is only one independent four-form, $\Psi^+$ or $\Psi^-$. 
This defines a $Spin(7)$-structure in the usual way. 

Although this case has been reviewed already in section \ref{oldstuff} 
let us check that one correctly recovers it from our equations. In taking the
limit one needs to be careful and consider only those equations obtained 
from spinor bilinears with four gamma matrices as these are the only equations
which are non-trivial. In fact, as written, the 
conditions 
on the $G_2$ structure in section \ref{bilinears} are, naively, all trivial in the
limit $\sin \zeta \rightarrow\pm1$. This is just because they are written in 
$G_2$-invariant form, whereas in this limit
there is no $G_2$ structure at all. An appropriate combination to consider 
is in fact
equation \reef{dY-eq} which we encountered in section \ref{cal}.
This reduces to the condition $\diff (\ex^{6\Delta} \Psi^\pm) = 0$
when $\sin \zeta \rightarrow\pm1$, and determines the internal space to be 
conformal to a $Spin(7)$ manifold, as in section \ref{oldstuff}. 
The electric flux reduces accordingly to 
\bea
G_{\mathrm{electric}} &  =  & \pm \vol_3 \wedge \diff (\ex^{3\Delta})~.
\eea
Notice that in fact we have set to zero only the irreducible $G_2$
components \rep{1} and \rep{7} of the magnetic flux, and in principle some
components are still allowed. Indeed, we recover the constraint on the
magnetic flux from equation
\reef{covariantK} which reduces to \reef{Tmn}, requiring the  flux to be 
in the \rep{27}$_+$ or \rep{27}$_-$ of $Spin(7)_\pm$, respectively.

Note that taking the $Spin(7)$ manifold to be a cone over a weak $G_2$
manifold and choosing the harmonic function $\ex^{-3\Delta}= 1/(mr)^6$
one again obtains AdS$_4\times M_7$ solutions, although now AdS$_4$
is foliated by $\bbR^{1,2}$ horospheres, with metric 
\bea
d\hat{s}^2_{11} & = & \ex^{-4ym} ds^2(\bbR^{1,2}) + dy^2 + d\tilde{s}^2_7~.
\eea

To summarise, we have shown that warped 
supersymmetric solutions with purely electric flux are of only two types:
the AdS$_3$ compactifications are in fact more naturally written as AdS$_4$
compactifications, foliated by copies of AdS$_3$, with the transverse space 
being weak $G_2$ holonomy. On the other hand in Minkowski$_3$ 
compactifications the internal manifold must be conformal to a 
$Spin(7)$-holonomy 
manifold, as discussed in \cite{Becker2},
with a single chiral spinor. Note that in the AdS$_3$ slicing case, the 
internal manifold $M_8$ provides a simple realisation of a space whose spinor
``interpolates'' between two spinors of opposite chirality.

\section{Examples}
\label{dyonicsection}

In this section we demonstrate that the formalism we have developed may be
useful for finding supersymmetric solutions. In particular, we easily
recover the dyonic M-brane solution of \cite{ILPT}. This describes a $1/2$-BPS 
M5/M2 bound state. We also argue that the 
recently discovered dyonic solutions of \cite{pope, warner} lie within this 
class,
although we will not attempt to rederive these solutions here. Indeed, all
of these solutions involve M5-branes with an M2-brane sitting inside. 
Finally, we present some simple solutions
to the equations of section \ref{puremagnetic}.

\subsection*{The dyonic M-brane}

As explained in section \ref{cal}, 
equation \reef{magnetic} is a generalised calibration condition
for an M5-brane wrapping an associative three-cycle in a $G_2$ manifold. 
Presently we shall regard $\bbT^3\oplus \bbR^4$ as a $G_2$ holonomy 
space\footnote{One may also consider the universal covering space $\bbR^7$, and wrap the brane over $\bbR^3$.} 
in which M5-branes wrap the three-torus $\bbT^3$. The remaining three
unwrapped world-volume directions span a $\bbR^{1,2}$ Minkowski space,
and we accordingly set $m=0$.   
Thus, it is natural
to write down the following simple metric ansatz describing such a wrapped 
brane:
\bea
d\hat{s}^2_{11} & = & \ex^{2\Delta}\left(ds^2(\bbR^{1,2}) + A\  d\mathbf{u}.d\mathbf{u} + H \ 
d\mathbf{x}.d\mathbf{x}\right)~.
\eea
Here $\mathbf{u}=(u_1, u_2, u_3)$ are coordinates on the three-torus and 
$\mathbf{x}=(x_1,\ldots, x_5)$ are coordinates on the Euclidean five-space  
transverse to the M5-brane. At this point $\Delta$, $A$ and $H$ 
are arbitrary functions on the internal eight-manifold. It is convenient
to choose the following orthonormal frame for the latter
\bea
e^{2+i} & = & A^{1/2} \diff u_{i} \nn
e^{5+\bar{a}} & = & H^{1/2} \diff x_{\bar{a}}
\eea
where $i=1,2,3$ and $\bar{a}=1,\ldots,5$. We then take the following $G_2$ 
structure on this eight-manifold
\bea
\phi & = & -e^{345} - e^3\wedge (e^{67}-e^{89}) - e^4\wedge (e^{68}+e^{79}) 
-e^5\wedge (e^{69}-e^{78})\nn
K & = & e^{10}~.
\eea
Thus we have written $\bbR^8 = \mathrm{Im}\bbH \oplus \bbH \oplus \bbR$, where
$\mathrm{Im}\bbH \oplus \bbH$ denotes the $G_2$-structure in its canonical
dimension of seven, and $\bbR$ is the $K$-direction. This appears to break
the invariance of the space transverse to the fivebrane under 
the five-dimensional Euclidean group, but in fact the solution we shall 
obtain respects this invariance -- it is simply not manifest in the above 
notation.

We now solve the equations of section \ref{bilinears}. 
Let us start with equation \reef{diff-K} for $K$ which is 
solved by taking
%
\bea 
\ex^{3\Delta} H^{1/2} \cos \zeta & = & c_1 
\eea
where $c_1$ is a constant. Equation \reef{starphi} gives the 
conditions
\bea
\ex^{6\Delta} A H &=& c_2^2 \\
\diff \left(\ex^{6\Delta}H^2\right) \wedge \diff x_{12345} &=& 0~.
\eea
One may solve the latter by taking $H=H(\mathbf{x})$, 
$\Delta=\Delta(\mathbf{x})$, which is natural as the solution 
should depend only on the coordinates transverse to the brane.
Using these relations one computes
\bea
A & = & \left(\frac{c_2}{c_1}\right)^2\cos^2 \zeta~.
\eea
Equation \reef{volume} is now automatically satisfied. One also computes
\bea
\diff \left(\ex^{6\Delta}\phi \cos \zeta\right) &=& \frac{c_2^3}{c_1} \diff 
u_{123} \wedge \diff\left(H^{-1}\cos^2 \zeta\right)
\eea
which implies that $\diff \phi \wedge \phi = 0$. Thus \reef{dphiwedgephi} gives
\bea
f  & = & 2 \sec \zeta \diff \zeta
\eea
and inserting this into the definition of $f$ \reef{electric} 
yields the following relation
\bea
H^{1/2}\tan \zeta & =&  c_4~.
\eea
We now set $c_2 = 1$ without loss of generality (by rescaling the coordinates 
$u_i$). The magnetic flux is obtained from \reef{magnetic} and reads
\bea
\ex^{3\Delta}F &= & -c_4 \,\diff u_{123}\wedge \diff (AH^{-1})+ c_1 \,\tilde{*}_5\diff H
\eea
where $\tilde{*}_5$ denotes the Hodge dual with respect to the metric 
$d\mathbf{x}.d\mathbf{x}$. Thus the Bianchi identity \reef{bianchiF} imposes
\bea
\tilde{\Box} H &=& 0~.
\eea
That is, $H$ is an harmonic function on the five flat transverse directions. 
One may easily 
check that the equation of motion \reef{f-equat} is identically satisfied.
It appears that we now have a solution with two free parameters, but this is
not so: one can remove $c_1$ by rescaling the coordinates $x_{\bar{a}}$. 
However, to recover\footnote{We disagree by factor of 6 with their 
expression for the flux. However, this appears to be a simple typographical 
error in taking the M-theory lift.} the solution of \cite{ILPT} we in
fact need to set
\bea
c_4 = -\tan \xi\qquad \qquad c_1=\cos \xi~.
\eea
We can choose $c_4 = -\tan \xi$ for some angle $\xi$ without loss of 
generality, and then setting
$c_1=\cos \xi$ corresponds to a specific choice of normalisation for the 
harmonic function. In conclusion, the metric takes the following form \cite{ILPT}
\bea
d\hat{s}^2_{11} = H^{-\frac{2}{3}}\left(\sin^2\xi + H 
\cos^2\xi\right)^{\frac{1}{3}}\left[
ds^2(\bbR^{1,2}) + \frac{H}{\sin^2\xi + H \cos^2\xi} 
d\mathbf{u}.d\mathbf{u} + H  
d\mathbf{x}.d\mathbf{x}\right]~.
\eea
Notice that the function $\zeta$ is given by
\bea
\tan^2\zeta & = & \frac{1}{H}\tan^2\xi
\eea
and that the M2-brane and M5-brane are recovered in the limits 
$\xi\rightarrow \pi/2$ and $\xi\rightarrow 0$, respectively.

Note that the solution actually preserves 16 Killing spinors 
\cite{ILPT}, as for the ordinary flat M5 brane. However, we have shown that 
the existence
of a $G_2$ structure of the type we have been discussing is enough
information to derive the full solution straightforwardly\footnote{By
a circle reduction to type IIA, followed by T-duality, one obtains
D-brane bound states in type IIB. The supersymmetry of the D5/D3 bound state 
\cite{BMM} is discussed in \cite{FG}.}.

\subsection*{``Dielectric flow'' solutions}

The solutions recently constructed in \cite{warnerold,pope,warner} fall
in our general class of ``dyonic'' solutions. Indeed they have a warped 
Minkowski$_3$ factor times an internal eight-manifold, and most importantly
have non-trivial electric and magnetic fluxes turned on. Thus they may be 
thought of
as some M5-brane distribution with induced space-filling M2-branes.  
Note that the solution of \cite{pope}, in particular, admits sixteen
supersymmetries -- as many as the dyonic M-brane of \cite{ILPT}.
In principle one should be able to recover these solutions 
in much the same way as we did for the standard dyonic M-brane solution 
above. All one has to do is to provide an ansatz for the
three-form $\phi$, or equivalently for the metric. Thus as shown in section
\ref{bilinears} the fluxes are determined by the supersymmetry constraints, and
one is left finally with a non-linear PDE to be solved.  Indeed, we have turned 
the problem into ``{\em algebraic}'' equations for the fluxes. While the solutions 
of \cite{ILPT,pope} preseve sixteen supercharges, and that of \cite{warner}
eight, our equations describe the most general ``dyonic'' solution, which
admits at least two Killing spinors with opposite chiralities. Thus these might
 be used to look for more general examples.

\subsection*{Smeared solutions}

Here we show that one may derive a simple class of solutions to
the equations of section \ref{puremagnetic}. One can think of these as
describing M5-branes wrapped on an associative three-cycle and
completely smeared over a $G_2$-manifold.
Unfortunately, these solutions are singular.
Of course, many of the singularities of supergravity solutions are ``resolved''
in M-theory. It would be interesting to know if this were the
 case here.

One makes the ansatz
\bea
\phi & = & \ex^{-3A(y)} \phi_0
\eea
where $\phi_0$ is the associative three-form for a $G_2$-holonomy manifold,
and assume in addition $\Delta=\Delta (y)$.
Thus, geometrically, we have a family of $G_2$-holonomy manifolds fibred
over the $y$-direction.
One finds that all of the differential equations for the structure are
satisfied automatically, apart from one, which imposes
\bea
\diff\left(\ex^{12\Delta}\vol_7\right)=0 \quad \Leftrightarrow \quad 
12\Delta(y) = 7A(y) + c~.
\eea
Notice that one may set $c=0$ by redefining $\phi_0$. Thus it remains to
 satisfy the Bianchi identity \reef{assPDE}. This imposes
\bea
\ex^{-6\Delta/7} & =  & a+by
\eea
where $a$ and $b$ are constants. Thus the solution is
\bea
d\hat{s}^2_{11}  & = & (a+by)^{-7/3} ds^2\left(\bbR^{1,2}\right) + 
(a+by)^{14/3}dy^2 + (a+by)^{5/3}ds^2(G_2)
\eea
where $ds^2(G_2)$ is any $G_2$-holonomy metric, and the $G$-flux is given
by
\bea
G & = & b (*_7\phi)_0
\eea
where $(*_7\phi)_0$ is the coassociative four-form on the $G_2$-manifold.
Setting $b=0$ gives $\bbR^{1,3}$ times a $G_2$-manifold. 
For $b\neq 0$ 
one may make a change of variables to write the metric as
\bea
d\hat{s}^2_{11} &  = &  dr^2 + r^{1/2}ds^2(G_2) + r^{-7/10}ds^2(\bbR^{1,2})~.
\eea
Clearly this is singular at $r=0$, although it is a perfectly regular 
supersymmetric solution everywhere else.

\section{Outlook}

In this paper we have studied the most general warped supersymmetric M-theory
geometry of the type $M_3\times M_8$, with the external space $M_3$ being
either Minkowski$_3$ or AdS$_3$. The key ingredient 
which allowed us to extend the analysis of \cite{Becker1,Becker2,AOG}
was to allow for an internal Killing spinor of indefinite chirality. This is 
in fact the 
most general form compatible with the three-eight decomposition and the 
Majorana
condition in eleven dimensions. The geometries were shown to admit a
particular $G_2$-structure. This is a special case of the most general 
eleven-dimensional geometry of the ``null'' type, for which the corresponding 
structure is $ ( Spin(7) \ltimes \bbR^8)\times \bbR$ \cite{M-wave,GP}. 

One of our motivations was to extend the analysis of \cite{Becker1,Becker2}
to more general supersymmetric geometries. However, it is a rather general 
result that, in the case of Minkowski$_3$ vacua, ignoring 
higher order corrections or singularities 
rules out compact solutions. We have noticed that
such corrections allow, in principle, compact geometries. It would be 
interesting to see if compact examples can be constructed.

We have found that the supersymmetry constraints also have a physical
interpretation in terms of generalised calibrations \cite{gkmw, GP, intrinsic}.
 In particular, 
we have shown most of the conditions arise as generalised
calibrations for ``dyonic'' M5-branes, namely M5-branes with M2-brane
charge induced on the world-volume by the three-form.
We have shown that when there is a suitable 
time-like Killing vector, one can construct a Bogomol'nyi bound in the 
presence of background $G$-flux. This applies for the entire class of 
geometries considered
here, and also to the ``time-like'' class of \cite{GP}. It would be 
interesting 
to understand more precisely the relation of generalised calibrations to the 
supersymmetry 
conditions in the general case of a $(Spin(7)\ltimes \bbR^8)\times \bbR$ 
structure, when the Killing vector is null.

The generality of our method implies that the conditions we have derived
apply to a variety of situations. Thus, apart from ``compactifcations'',
one can use the same results to describe non-compact geometries of physical
interest. Typical examples are wrapped branes or intersecting branes.
In these, as in all other cases, the supersymmetry constraints are relatively
easy to implement, while ensuring that the Bianchi identity is satisfied is 
often a challenging task. One generically obtains non-linear PDEs whose 
explicit solutions 
are typically beyond reach. In any case, as illustrated in section
\ref{puremagnetic}, it should be clear that our approach is suitable for
generalising the work of \cite{FS}. In particular, we have given the 
conditions
and PDEs describing M5-branes wrapped on associative and SLAG three-cycles.
In the last case one can show that the Calabi-Yau three-fold becomes a 
non-Hermitian manifold after allowing for the backreaction.
This is to be contrasted with the case where M5-branes (or NS5-branes 
in type II) wrap holomorphic cycles. Here the holomorphic structure
of the manifold is preserved
\cite{hulltorsion,strominger,FS,FSN=1,intrinsic,ALE}.

Rewriting the Killing spinor equations in terms of the underlying $G$-structure
provides an elegant organisational principle, and sheds light on the 
geometry of supersymmetric solutions. Namely, it turns out 
that
the geometrical interpretation of the fluxes is given by the intrinsic torsion.
Much physical insight comes from the interpretation of these in terms
of branes and calibrations. On the other hand, the complication that arises
from solving the equations implied by the Bianchi identitity seems to be a
limitation on the method for finding new solutions. It is conceivable that 
using the geometrical and physical insights of our approach in combination 
with other techniques, such as those related to gauged supergravities, will
improve the situation. Some ideas in this direction have already appeared
(see, {\it e.g.}, \cite{warner}) and it would be interesting to elaborate on 
them further.

\subsection*{Acknowledgments}
We would like to thank Jerome Gauntlett and Dan Waldram for very useful discussions.
D.~M.~ is supported by an EC Marie Curie Individual Fellowship under contract 
number HPMF-CT-2002-01539. J.~S.~ is supported by an EPSRC mathematics 
fellowship.

\appendix

\section{$G_2$-structures}
\label{g2appendix}

A $G_2$-structure on a seven-dimensional manifold is specified by an 
associative three-form $\phi$, which in a local frame may be written 
\begin{equation}
   \phi = e^{246}-e^{235} - e^{145} - e^{136} 
      + e^{127} + e^{347} + e^{567}~ .
      \label{g2form}
\end{equation}
This defines uniquely a metric $g_7=(e^1)^2+\dots+(e^7)^2$ and an orientation
$\vol_7=e^1\wedge\dots\wedge e^7$. We then have 
\begin{equation}
   *\phi = e^{1234}+e^{1256}+e^{3456}+ e^{1357}-e^{1467}-e^{2367}-e^{2457}~.
\end{equation}
The adjoint representation of $SO(7)$ decomposes as
\rep{21}$\to$\rep{7}+\rep{14} where \rep{14} is the adjoint
representation of $G_2$. We therefore have $g_2^\perp\simeq$ \rep{7}.
The intrinsic torsion then decomposes into four modules~\cite{fernandez}: 
\begin{equation}
\begin{aligned}
   T \in \Lambda^1 \otimes g_2^\perp 
      &= \mathcal{W}_1 \oplus \mathcal{W}_2 \oplus \mathcal{W}_3
         \oplus \mathcal{W}_4 , \\
   \mathbf{7} \times \mathbf{7} 
      &\to \mathbf{1}  + \mathbf{14} + \mathbf{27}+ \mathbf{7}~ .
\end{aligned}
\end{equation}
The components of $T$ in each module $\mathcal{W}_i$ are encoded in
terms of $\diff\phi$ and $\diff *\phi$ which decompose as 
\begin{equation}
\begin{aligned}
   \diff \phi \in \Lambda^4 
      &\cong \mathcal{W}_1 \oplus \mathcal{W}_3 \oplus \mathcal{W}_4  \\
   \mathbf{35} &\to \mathbf{1}+ \mathbf{27} + \mathbf{7}   \\
   \diff * \phi \in \Lambda^5 
      &\cong \mathcal{W}_2 \oplus \mathcal{W}_4  \\
   \mathbf{21} &\to \mathbf{14} + \mathbf{7} ~.
\end{aligned}
\end{equation}
Note that the $W_4$ component in the \rep{7} representation appears
in both $\diff\phi$ and $\diff *\phi$. It is the Lee form, given by
\begin{equation}
   W_4 \equiv \phi \lrcorner\, \diff \phi = -*\phi\lrcorner\, \diff *\phi.
\end{equation}
The  $\mathcal{W}_1$ component in the singlet representation can be
written as 
\begin{equation}
   W_1 \equiv *(\phi \wedge \diff \phi) .
\end{equation}
The remaining components of $\diff \phi$ and $\diff *\phi$ encode $W_3$ and
$W_2$, respectively. The $G_2$ manifold has $G_2$ holonomy if and only if
the intrinsic torsion vanishes, which is equivalent to $\diff \phi=\diff *\phi=0$. 
Note that $G_2$-structures of the type ${\cal W}_1\oplus{\cal
W}_3\oplus{\cal W}_4$ are called integrable as one can introduce a $G_2$
Dolbeault cohomology \cite{fernug}.

On a manifold with a $G_2$-structure forms decompose into irreducible $G_2$
represenations. In particular, we have the following decompositions
of the spaces of two-forms and three-forms:
\begin{equation}
\begin{aligned}
   \Lambda^2  
      &= \Lambda^2_7 \oplus \Lambda^2_{14}\\
   \Lambda^3 
      &=\Lambda^3_1 \oplus \Lambda^3_7 \oplus \Lambda^2_{27} ~.\\
\end{aligned}
\end{equation}
The Hodge dual spaces $\Lambda^5$ and $\Lambda^4$ decompose accordingly.
For applications in the main part of the paper, it is useful to write down 
explicitly the decompositions of the three-forms and four-forms. A three-form
$\Omega \in \Lambda^3$ is decomposed into $G_2$ irreducible representations as
\bea
\Omega & = & \pi_1 (\Omega) + \pi_7 (\Omega)  + \pi_{27} (\Omega)
\eea
where the projections are given by
\bea
\pi_1 (\Omega) & = & \frac{1}{7} (\Omega \con \phi) \, \phi\nn
\pi_7 (\Omega) & = & -\frac{1}{4} (\Omega \con *\phi)\con * \phi\nn
\pi_{27} (\Omega)_{ijk} & = & \frac{3}{2} \hat{Q}_{r[i}\phi^r{}_{jk]}
\eea
and $\hat{Q}_{ij}$ is the traceless symmetric part of the tensor
$Q_{ij}=\frac{1}{2!}\Omega_{ikr}\phi^{kr}{}_j$, namely
\bea 
Q_{ij} & = & \frac{3}{7} (\Omega \con \phi ) g_{ij} - \frac{1}{2} \phi_{ij}{}^k
(\Omega \con *\phi)_k + \hat{Q}_{ij}~.
\eea
Similarly, a four-form $\Xi\in \Lambda^4$
decomposes into $G_2$ irreducible represenations as 
\bea
\Xi & = & \pi_1 (\Xi) + \pi_7 (\Xi)  + \pi_{27} (\Xi)
\eea
where the preojections are given by
\bea
\pi_1 (\Xi) & = & \frac{1}{7} (\Xi \con *\phi) *\phi\nn
\pi_7 (\Xi) & = & -\frac{1}{4} (\phi \con \Xi)\wedge \phi \nn
\pi_{27} (\Xi)_{ijkm} & = & -2 \hat{U}_{r[i} *\phi^r{}_{jkm]}
\eea
and $\hat{U}_{ij}$ is the traceless symmetric part of the tensor
$U_{ij}=\frac{1}{3!}\Xi_{ikrm} *\phi^{krm}{}_j$, namely
\bea 
U_{ij} & = & -\frac{4}{7} (\Xi \con * \phi ) g_{ij} - \frac{1}{2} \phi_{ij}{}^k
(\phi \con \Xi)_k + \hat{U}_{ij}~.
\eea

Consider an infinitesimal variation of the associative three-form 
$\delta \phi $ and the induced variations of the metric 
$\delta g_{ij}$,  and coassociative four-form $\delta *\phi $. Using the 
various identities obeyed by the  $G_2$ structure, 
we obtain an explicit decomposition of $\delta \phi$, namely
\bea
\pi_1 (\delta \phi) & = & \frac{3}{7} \delta \log\sqrt{g} \,\phi\nn
\pi_7 (\delta \phi) & = & - \frac{1}{4}(\delta \phi \con *\phi) \con * \phi\nn
\pi_{27} (\delta \phi)_{ijk} & = & \frac{3}{2} \delta g_{r[i}\phi^r{}_{jk]}
-\frac{3}{7} \delta \log\sqrt{g}\, \phi_{ijk}~.
\label{dphiproj1}
\eea 
The irreducible components of $\delta *\phi$ are similarly given by
\bea
\pi_1 (\delta * \phi) & = & \frac{4}{7} \delta \log\sqrt{g} *\phi\nn
\pi_7 (\delta * \phi) & = & -\frac{1}{4} (\phi \con \delta *\phi) \wedge \phi\nn
\pi_{27}(\delta * \phi)_{ijkm} & = & 2 \delta g_{r[i} *\phi^r{}_{jkm]}
-\frac{4}{7} \delta \log\sqrt{g}\, * \phi_{ijkm}~.
\label{dphiproj2}
\eea
The following relations also hold
\bea
\frac{1}{2!}\delta\phi_{(i|kr} \phi^{kr}{}_{j)} & = & \delta g_{ij} + g_{ij}
\delta \log\sqrt{g}\nn
\frac{1}{3!}\delta * \phi_{(i|krm}  * \phi^{krm}{}_{j)} & = & - \delta g_{ij} -2
g_{ij}\delta \log\sqrt{g}\nn
\phi \con \delta *\phi & = & - \delta \phi \con *\phi\nn
\pi_{27} (\delta * \phi ) & = & - * \pi_{27} (\delta \phi)~.
\label{dphiproj3}
\eea
Using these expressions one can derive the following useful
equation
\bea
\delta * \phi & = & - * \delta \phi  + \delta\log\sqrt{g} *\phi 
+\frac{1}{2} (\delta \phi \con * \phi)\wedge \phi~.
\label{nice}
\eea

\section{The M5-brane Hamiltonian}
\label{m5}

In this appendix we present a brief discussion of the Hamiltonian formulation 
of the M5-brane world-volume theory \cite{townsend}. We use this to obtain
an expression for the energy of a class of static M5-branes, which, in the
main text, is shown to satisfy a Bogomol'nyi-type inequality. We also 
recall some details of the M5-brane $\kappa$-symmetry.

The action of the M5-brane is complicated by the presence of a self-dual 
three-form $H$ which propagates on the world-volume. This requires one to 
introduce an auxilliary scalar field $a$ (see \cite{Sorokin} for a review),
with a normalised ``field strength'' $v_i=\partial_i a/\sqrt{-(
\partial a)^2}$. One then has an additional gauge invariance that one may
use to gauge-fix $a$, at the expense of losing
manifest spacetime covariance. However, the Hamiltonian treatment requires 
one to make 
a choice of time coordinate. 
Using the symmetries of the M5-brane action, one may then choose the 
``temporal gauge'' $a = \sigma^0 = t$, where $\sigma^i =(\sigma^0,\sigma^a)$
are world-volume coordinates ($a=1,\ldots, 5$), and the background 
spacetime is assumed to take the static form $d\hat{s}_{11}^2=-\ex^{2\Delta}
dt^2+ds^2_{10}$. One then proceeds with the Hamiltonian approach 
\cite{townsend}, which yields the constraints
\bea
\tilde{P}^2+T_{M_5}^2 L_{DBI}^2 & = & 0~{}\nn
\partial_aX^M \tilde{P}_M  & = & 0~.\label{ham}
\eea
Here $X^M=(t,X^I)$ are the embedding coordinates, $T_{M_5}$ is the M5-brane 
tension,  $L_{DBI} = 
\sqrt{\det\left(\delta_a^{\ b} + H^{* b}_a\right)}$ is a Born-Infeld-like
term, and 
\bea
\tilde{P}^M & = & P^M + T_{M_5}\left(V^a \partial_a X^M - \mathcal{C}^M\right)~.
\eea
We have that 
\bea
V_c = \frac{1}{4} H^{*ab}H_{abc}
\eea
where the two-form $H^* = *_5 H$ is the \emph{world-space} dual of $H$ 
(the $H_{0ab}$
components of $H$ will not contribute to the energy) and  
the term $\mathcal{C}_M$ is a contribution
from the Wess-Zumino couplings of the M5-brane, namely
\bea
\mathcal{C}_M  & = & *_5\left[i _M C_6 -\frac{1}{2} i_M C \wedge
(C-2H)\right]
\eea
where $i_M$ denotes interior contraction with the vector field $\partial/
\partial X^M$. Recall that the Wess-Zumino coupling 
of the M5-brane is given by
\bea
I_{WZ} & = & \int_W C_6 + \frac{1}{2}C\wedge H
\label{WZ}
\eea
where $H$ is the three-form field strength on the five-brane, coupled to the
background $C$-field
\bea
H = h + j^*C~.
\eea
Here $h$ is closed, and locally of the form $h=\diff\, b$ for some two-form 
potential $b$. 
Notice that $\diff H = j^*G$, where $j$ is the M5-brane embedding map.

We may now use the Hamiltonian and momentum constraints \reef{ham} to 
obtain an expression for the energy density. We consider static configurations
with $\tilde{P}^I=0$. This is sufficient to satisfy the momentum constraint,
but not in general necessary. One could extend our analysis to the general
case (with more effort), but we will not do this here -- the class of static
configurations we consider will be general enough for our purposes.
One defines the energy in the usual way
\bea
E   =  -P^M k^N \hat{g}_{MN} = -P_0 = \ex^{2\Delta}P^0
\eea
where $k$ is the time-like Killing vector field $\partial/\partial t$. The
Hamiltonian constraint now allows one to solve for the energy
\bea
E & = &  T_{M_5}\left(\mathcal{C}_0 + \ex^{\Delta}L_{DBI}\right)~.
\label{energy}
\eea

In addition to the energy, the other ingredient we use in the main text  
is the $\kappa$-symmetry and supersymmetry transformations of the fermions. 
These combine to give
\bea
\delta\theta  & = & \mathcal{P}_+\kappa + \eta\label{kappa}
\eea
where $\mathcal{P}_{\pm} = 
\frac{1}{2}(1\pm\tilde{\Gamma})$ are projector operators. 
$\eta$ is the background 
supersymmetry $Spin(1,10)$ spinor, and $\tilde{\Gamma}$ is a traceless
Hermitian product structure, that is, $\mathrm{tr}\ \tilde{\Gamma}=0$, 
$\tilde{\Gamma}^2=1$, $\tilde{\Gamma}^{\dag}=\tilde{\Gamma}$. 
Explicitly, we have
\bea
\tilde{\Gamma}  & =  & \frac{1}{L_{DBI}}
\ex^{-\Delta}\hat{\Gamma}_0\left[V\cdot\tilde{\gamma}+\frac{1}{2}
\tilde{\gamma}^{ab} H^*_{ab}
+\frac{1}{5!}\tilde{\gamma}_{a_1\ldots a_5}\epsilon^{a_1 \ldots a_5}\right]
\eea
where $\tilde{\gamma}^a$ are the pull-backs of the eleven-dimensional 
Clifford matrices to the M5-brane world-space. If we consider static 
configurations with a rest frame that has zero momentum, then $V_a = 0$.
This is the form of the projector used in the main text. One can show 
\cite{BKOP} that the variation \reef{kappa} vanishes if, and only if,
\bea
\mathcal{P}_-\eta & = &  0
\eea
which therefore characterises bosonic supersymmetric configurations.

\section{Useful relations}
\label{relations}

Given the supersymmetry equations 
\reef{dilatino-like}, and using the symmetry properties of the gamma matrices, 
one can derive some useful identities which we have used extensively 
in deriving
our results. For the reader's convenience we list them here:
\bea
\frac{1}{288} F_{pqrs} \e^{\pm \trsp} [\gamma^{pqrs},A]_- \e^\pm \mp
\frac{1}{2}\de_m \Delta \e^{\pm\trsp} [\gamma^{m},A]_- \e^\pm + 
m (\e^{\mp \trsp}A\e^\pm - \e^{\pm \trsp}A \e^\mp) \nn
\mp \frac{1}{6}f_m \e^{\pm \trsp} A \gamma^m \e^{\mp} 
\pm \frac{1}{6}f_m \e^{\mp \trsp}\gamma^m A \e^{\pm} = 0\\
\frac{1}{288} F_{pqrs} \e^{\pm \trsp} [\gamma^{pqrs},A]_+ \e^\pm \mp
\frac{1}{2}\de_m \Delta \e^{\pm\trsp} [\gamma^{m},A]_+\e^\pm + 
m (\e^{\mp \trsp}A\e^\pm + \e^{\pm \trsp}A \e^\mp) \nn
\pm \frac{1}{6}f_m \e^{\pm \trsp} A \gamma^m \e^{\mp} 
\pm \frac{1}{6}f_m \e^{\mp \trsp}\gamma^m A \e^{\pm} = 0\\
\frac{1}{288} F_{pqrs} \e^{+ \trsp} [\gamma^{pqrs},A]_{\pm} \e^- 
-\frac{1}{2}\de_m \Delta \e^{+ \trsp} [\gamma^{m},A]_{\mp}\e^- + 
m (\e^{-\trsp}A\e^- \pm \e^{+\trsp}A \e^+) \nn
\mp \frac{1}{6}f_m \e^{+ \trsp} A \gamma^m \e^{+} 
+ \frac{1}{6}f_m \e^{- \trsp}\gamma^m A \e^{-} = 0
\eea
where $[\cdot,\cdot]_\pm$ refers to an anticommutator or commutator, and $A$ is
a general Clifford matrix.

\end{document}